\documentclass[10pt, conference]{IEEEtran}
\usepackage{amssymb}
\usepackage{amsmath}
	\renewcommand{\theequation}{\arabic{equation}}
\usepackage{graphicx}
\usepackage{enumitem}
\usepackage{url}
\usepackage{ifthen}
\usepackage{xspace}
\usepackage{xfrac}
\usepackage{float}
\usepackage[tight,footnotesize]{subfigure}
\usepackage{array}
\usepackage{xfrac}
\usepackage{fancyvrb}
\usepackage{multirow}

\usepackage{booktabs}
\usepackage[table]{xcolor}
\definecolor{lightGray}{gray}{0.85}

\usepackage[framemethod=TikZ]{mdframed}
\usepackage{lipsum}
\mdfdefinestyle{MyFrame}{%
    linecolor=gray!90!white, 
    outerlinewidth=1.5pt,
    roundcorner=5pt,
    innertopmargin=0.2\baselineskip,
    innerbottommargin=0.2\baselineskip,
    innerrightmargin=3pt,
    innerleftmargin=3pt,
    backgroundcolor=white}

\mdfdefinestyle{MyWhiteBlackFrame}{%
    linecolor=gray!90!white, 
    outerlinewidth=0.5pt,
    roundcorner=5pt,
    innertopmargin=0.2\baselineskip,
    innerbottommargin=0.2\baselineskip,
    innerrightmargin=2pt,
    innerleftmargin=2pt,
    backgroundcolor=white}

\newboolean{showcomments}
\setboolean{showcomments}{false}
\ifthenelse{\boolean{showcomments}}{%
   \newcommand{\bnote}[2]{
    \marginpar{\fbox{\bfseries\sffamily #1}}
    {\sffamily\small{\large $\P$}#2{\large $\P$}}}
}{%
   \newcommand{\bnote}[2]{}
}

\newcommand{\ie}{i.e.,\xspace}
\newcommand{\eg}{e.g.,\xspace}
\newcommand{\etal}{\emph{et al.}\xspace}
\newcommand{\etc}{etc.\xspace}

\newcommand{\secref}[1]{Section~\ref{#1}\xspace}
\let\Secref\secref

\let\Figref\figref

\let\Tabref\tabref

\newcounter{rq}
\newcounter{q}
\newenvironment{Marking}{
	\vspace{+0.1cm}
	\begin{mdframed}[style=MyWhiteBlackFrame]
}
	{\end{mdframed}}

\newcounter{f}
\newcounter{deflist}
\newcounter{mydef}
\newcommand{\myparagraph}[1]{\noindent\textbf{#1}}

\begin{document}

\title{Modeling and Analyzing Release Trajectory based on the Process of Issue Tracking}

\author{
	\IEEEauthorblockN{
		Hani Abdeen and
		Houari Sahraoui
		}
	\IEEEauthorblockA{
		DIRO,
		Universit\'e de Montr\'eal, Montr\'eal(QC),
		Canada\\
		\{abdeenha,sahraoui\}@iro.umontreal.ca}
	}

\maketitle

\begin{abstract}
Software release development process, that we refer to as \emph{``release trajectory''}, involves development activities that are usually sorted in different categories, such as incorporating new features, improving software, or fixing bugs, and associated to ``issues''.
Release trajectory management is a difficult and crucial task.
Managers must be aware of every aspect of the development process for managing the software-related  issues.
Issue Tracking Systems (ITSs) play a central role in supporting the management of release trajectory.
These systems, which support reporting and tracking issues of different kinds (such as ``bug'', ``feature'', ``improvement'', \etc), record rich data about the software development process.
Yet, recorded historical data in ITSs are still not well-modeled for supporting practical needs of release trajectory management.

In this paper, we describe a sequence analysis approach for modeling and analyzing releases' trajectories, using the tracking process of reported issues.
Release trajectory analysis is based on the categories of tracked issues and their temporal changing, and aims to address important questions regarding the co-habitation of unresolved issues, the transitions between different statuses in release trajectory, the recurrent patterns of release trajectories, and the properties of a release trajectory.
\end{abstract}

	
\IEEEpeerreviewmaketitle

\vspace{-0.2cm}
\section{Introduction}\label{sec:introduction}
\vspace{-0.1cm}
Software release trajectory is part of the software evolution process that leads from the software initial development, or from a released version, to a new release available to the software users \cite{Hoek97, Blok2008}. It includes the development activities such as adding new features, improving software, and fixing bugs. These activities change the software state \cite{Hind07}. For example, fixing a bug B changes the software state from ``defected by B'' to ``cleaned of B'', and fixing all identified bugs changes the software state from ``defected'' to ``clean''.

Release managers have the responsibility to control the release schedule and document the release state.
Therefore, they need supporting tools to help them to understand the release trajectory and to improve their process planning and control skills. In this work, we are interested in techniques that can characterize the process of development activities that lead to a software release, and help software managers to better understand the evolution of software state through the software release.
Unlike existing work on understanding software evolution and code changes (\eg \cite{Hind09a, Koth06, Xing:2004:1357808, Barry2003, Kemerer1999}), we seek an approach that describes the evolution of software state from release to release based on concrete diagnostics reported by the software managers. 
The approach should help software managers in understanding the software state in the beginning of a release trajectory (\eg the software was suffering some bugs), and what transitions were achieved to reach the state which demarcates the underlined release.

Nowadays, Issue Tracking Systems (ITSs) play a central role in supporting software-development  management \cite{Baysal13}.
ITSs are used by software developers and managers to report, track and manage issues of different kinds (such as ``bug'', ``new feature'', ``improvement'', ``technical task'', ``test'', ``documentation'' \etc).
Recorded issues in ITSs summarize the development activities that lead to software release, and the types of recorded issues represent the categories of development activities.
Indeed, an opened, and accepted, issue of type ``bug'' means that the software suffers from the underlined bug. 
When this bug issue is closed and recorded as fixed, this means that the current release has sorted out a bug by performing specific development activity relevant to the bug fix.
Hence, release notes are usually generated from ITSs recorded data, which include information about the fixed/implemented issues and their types \cite{More14}.


Nevertheless, recorded historical data in ITSs are modeled and presented in a way that does not support adequately  practical needs of release trajectory comprehension. 
Actually, well known ITSs, such as JIRA, provide helpful tools for filtering and sorting issues according to different criteria. 
For instance, users can list all resolved issues within a period of time and sort them according to their closing dates.
Despite these facilities, it is not trivial to retrieve the evolution process of issues tracking in ITSs.
For instance, managers may be interested in addressing the following questions:
what type of issues was closed first, ``bug'', ``feature'' or ``improvement''?
and what type was closed next?
when and for how long time ``improvement'' issues were unresolved (\ie open)?
what type(s) of unresolved issues co-existed together in a given period?
when and how this state was changed?

\begin{figure*}[t!]
\centering
	\includegraphics[width=0.95\linewidth]{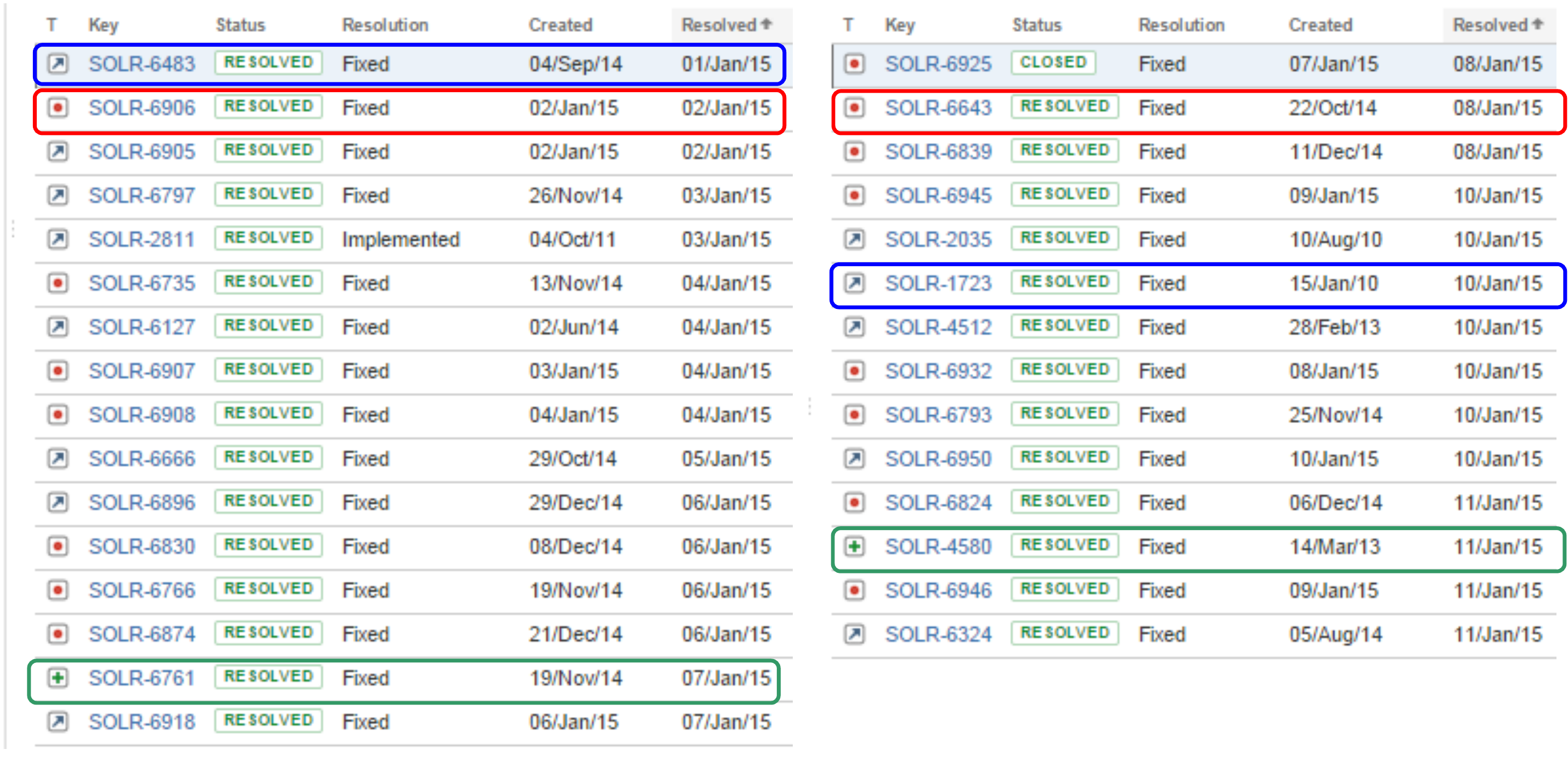}
	
%
\vspace{-0.3cm}
\caption{List of resolved (fixed) issues (of types ``bug'', ``improvement'', or ``new feature'') in Apache-Solr project between 1-Jan-2015 and 11-Jan-2015, as recorded in JIRA ({http://issues.apache.org/jira/issues}).
Issues are sorted from first resolved issue (top-left) to latest resolved issue (bottom-right).
}
\label{fig:SOLR_Issues_Example}
\end{figure*}

For example, \Figref{fig:SOLR_Issues_Example} shows the list of resolved (and fixed) issues for the project Apache-Solr \cite{Solr} in a period of 11 days, between 1-Jan-2015 and 11-Jan-2015.
This list includes only 30 issues that are sorted from first to last resolved issue.
Users can easily observe that the improvement issue SOLR-6483 (in the top left line) was resolved before the bug issue SOLR-6906 (in the second line), as well as before the improvement issue SOLR-1723.
However,  it is not trivial to observe that, according to this list, improvement issues were open since 15-Jan-2010 (the creation date of issue SOLR-1723) and the life-cycle of improvement queries span without interruptions until 11-Jan-2015 (when the issue SOLR-6324, in the bottom right line, was closed).
Indeed, during all this period, there was at least one improvement issue unresolved yet.
In the same vein, it is not trivial to observe that bug fixing queries span the period from 22-Oct-2014 (the creation date of issue SOLR-6643) to 11-Jan-2015 (when the issue SOLR-6946  was closed), also without interruptions. Additionally, queries for new features span the period from 14-Mar-2013 to 11-Jan-2015 (the creation and closing dates of the issue SOLR-4580).

In this paper, we propose a sequence analysis approach for modeling and analyzing the releases' trajectories using the tracking process of reported issues in ITSs.  
It employs sequence-analysis techniques for analyzing releases' trajectories, generating a new kind of reports summarizing the trajectories and mining recurrent patterns. 
Our approach aims to address important questions regarding the co-habitation of unresolved issues, the transitions between different software statuses in release trajectory, the recurrent patterns of release trajectories, and the properties of a release trajectory.


To evaluate our approach, we performed a descriptive study of releases' trajectories using a fairly large dataset of releases, 84 releases, that we extracted from the historical evolution of three large software projects. Our study shows that our approach contribute to the understanding of the development processes of releases providing new kinds of reports that can assist managers in better documenting releases, in auditing the behavior of releases development (that is implicitly the behavior of involved software developers), and in identifying recurrent patterns of development paths of releases.   

This paper is organized as follows. Sections \ref{sec:background} and \ref{sec:approach} present respectively the background and the approach for modeling and analyzing releases' trajectories. 
The evaluation of our approach is presented in \Secref{sec:evaluation}.  
Finally, \Secref{sec:relatedWork} present the relevant existing work before concluding in \Secref{sec:conclusion}.

\section{Background}\label{sec:background}
Past studies of software evolution (such as \cite{Kemerer1999, Herraiz2009, Hind09b, Lin2013, Goul12a}) widely employed time-series analysis to study historical software change events.
In time-series analysis, continuous data represented as successive data points are required, and techniques such as ARIMA (autoregressive integrated moving average) are used to fit models to time-series data to allow a deeper understanding of data evolution \cite{Kemerer1999, Goul12a}.
To apply time-series analysis, data points must be uniformly spaced, and hence a suitable time unit (period length) must be specified for obtaining sufficiently stationary time series.

An alternative to time-series analysis is the sequence analysis (also known as trajectory analysis), which is the statistical study of succession of states or events that are usually ordered temporally\footnote{A sequence may also reflect other types of order, such as spatial order, preference order, hierarchical order, \etc} \cite{sequence-analysis14}. 
Unlike time-series analysis, sequence analysis is based on analysis of categorical data, rather than continuous data.
It has been widely used in social science to analyze and understand biographical trajectories, trajectories of cohabitation, housing, career, \etc
Sequence analysis aims at identifying sequences of states in the developmental process, where each individual is characterized by his unique trajectory.
A trajectory is defined as a string of states with specific order and each state has a specific duration.
Similarities between trajectories can then be measured based on shared sub-sequences.
Sequence analysis can address questions about the types of developmental processes, the structural properties of a process, and the factors influencing different types of
development paths.

In this paper, we are interested in learning how a software state evolves over time through a software release based upon software tracked issues and the temporal changing of their status (from open/unresolved to closed/resolved). 
More specifically, we are interested in: 
1) understanding whether development processes of software releases follow similar trajectories; 
2) identifying patterns in releases' trajectories;  
3) and identifying frequent and consistent transitions between distinct states in releases' trajectories.  


\section{Release Trajectory Analysis Approach}\label{sec:approach}
In this section, we describe our sequence-analysis approach for modeling and analyzing release trajectory based on the evolution process of software state.
Our approach relies on the software diagnostics as reported by the software managers and developers in the form of issues, and on the tracking process of reported issues in the issue tracking system.

By definition the trajectory of a given software release, R, is the evolution process of the software state starting from the first moment registered in the release development process, the {\it inception timestamp} of R, and ending by the moment that demarcates the development iteration leading to the targeted release, the {\it ending timestamp} of R.
According to this definition, a release is recognized by its identifier (id) and its inception and ending timestamps, and characterized by the software states (as well as the transitions between states) that occurred between these timestamps. Hence, a release is concerned only by the resolved issues (\ie the performed changes) between those timestamps \cite{More14}.


In the remainder of this section, we describe the required data before detailing our approach for modeling and analyzing software release trajectory.

\begin{figure}[!t]
\centering
	\includegraphics[width=0.8\linewidth]{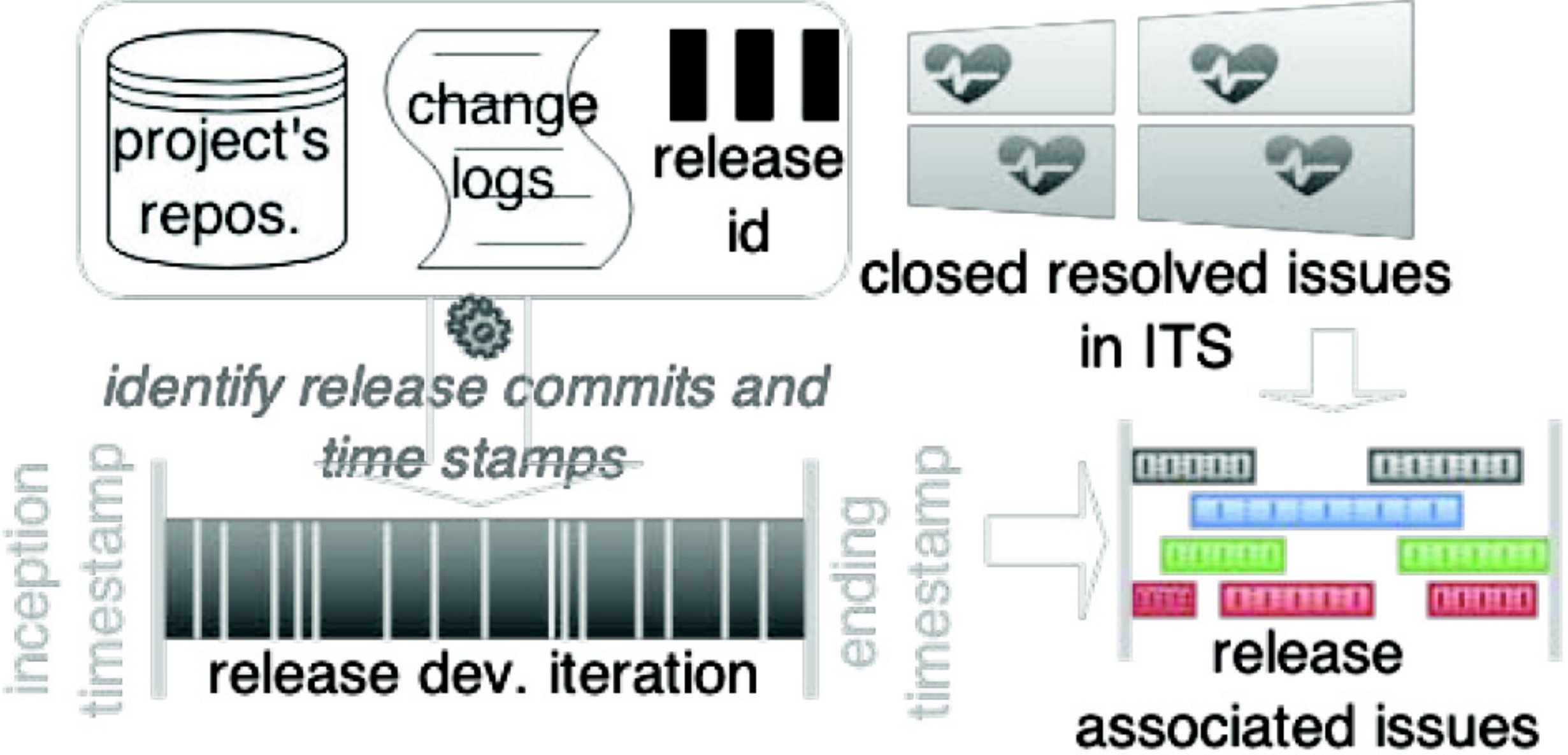}
\vspace{-0.2cm}
\caption{Overview of required data by our approach.}
\label{fig:Data}
\end{figure}

\subsection{Required Data}\label{sec:requiredData}
\Figref{fig:Data} provides an overview of required data to apply our approach.
First of all, for analyzing the trajectory of a given software release, the inception and ending timestamps of the underlined release must be provided.
By definition, these timestamps are respectively the timestamps of the start (earliest) and end (latest) commit revisions for the underlined release.
Therefore, we need to access the change log of the studied system and retrieve the commits on both the trunk and branches.
Unfortunately, identifying the earliest and latest commits for releases usually requires human intervention \cite{Shob14}.
For certain software projects these commits can be identified easily thanks to some standard tags in the commit log, that are generated automatically by release preparation tools.
For instance, when Maven release plugin is used to release a project, this automatically inserts two commits in the project change log with specific log messages:
{\it ``$[$maven-release-plugin$]$ prepare release \texttt{RELEASE\_ID}''} (this is the latest commit for the release \texttt{RELEASE\_ID}),
{\it ``$[$maven-release-plugin$]$ prepare for next development iteration''} (this is the earliest commit for the next release).
If commit logs do not provide such specific tags, we need to map commits for releases using the approach of Shobe \etal \cite{Shob14}. Thus, we need to determine the names of releases and their representative branches in the project's repository, as described in \cite{Shob14}.

To retrieve the release trajectory, our approach relies on the reported issues in the software ITS that are resolved, closed, and associated to the considered release.
Here, we consider that an issue is ``resolved'' when it was stated explicitly as treated and resolved --\ie the issue resolution parameter is explicitly set to ``fixed'', ``implemented'', or other similar terms; but not to ``invalid'', ``not a problem'', ``not complete'', or other similar terms.
For some closed issues, we may find that developers have manually associated them to the concerned release, but unfortunately this is not always the case.
Therefore, we select all issues that were closed and resolved during the development iteration of the considered release, that is determined by the inception and ending timestamps of the release.


After identifying the resolved issues, issues can be filtered according to their types as ``included'' or ``excluded'', and included issues can be sorted as ``recurrent (relevant)'' or ``non-recurrent (not relevant)''.
In this work, all standard issue types\footnote{In this paper we mainly use the terminology of the Atlassian JIRA when we refer to issues types, categories, parameters.} are included,
\eg ``Bug'', ``Dependency Upgrade'', ``Documentation'', ``new Feature'', ``Improvement'', ``Task'', ``Test'', ``Wish'' \etc, whilst the ``sub-task'' issue type is excluded.


Note that the list of standard issue types may vary from one project to another, and it could be very long. However, only a small subset of the aforementioned types can be considered as recurrently reported in different software projects and reported in release notes. These are:
``Bug'', ``new Feature'', ``Improvement'', and ``Task''.
The first three types have software development story nature
(\eg
Bug issue:
{\it ``Unhandled exception thrown from within ...''};
Improvement issue:
{\it ``Reduce object allocations''};
new Feature issue:
{\it ``Add a policy interface for targeting ...''}
),
whilst Task issues have technical nature (\eg Task issue: {\it ``Upgrade to latest JPA 2.0 TCK''}).
Recurrent issue types are processed by our approach separately and modeled explicitly in the releases' trajectory, whilst non-recurrent types, such ``Documentation'' and ``Wish'', are aggregated all together in one virtual type.


Furthermore, in this work, we assume that sub-task issues should be excluded as they are related to other issues having standard types.
More precisely, as the name ``sub-task type'' indicates, sub-task issues are useful for splitting up a parent issue having a standard type into a number of smaller tasks that can be assigned and tracked separately\footnote{Refer to JIRA documentation: https://confluence.atlassian.com/display/ JIRA/Creating+a+Sub-Task}.
Hence, a resolved sub-task issue means that only a partial development activity has been achieved regarding the parent issue. We assume that the software state does not change until resolving the parent issue itself.

Once the aforementioned data about releases and issues are collected, our approach for modeling releases' trajectories can be applied.
Additionally, when commit logs are available and associated to (tagged by) the recorded issues, our approach can use commits to provide further insights on a release trajectory from commits-activity perspective.

\vspace{+0.9cm}
\subsection{Release Trajectory Modelling}\label{sec:trajectoryModelling}

\begin{figure}[!t]
\centering
	\includegraphics[width=0.95\linewidth]{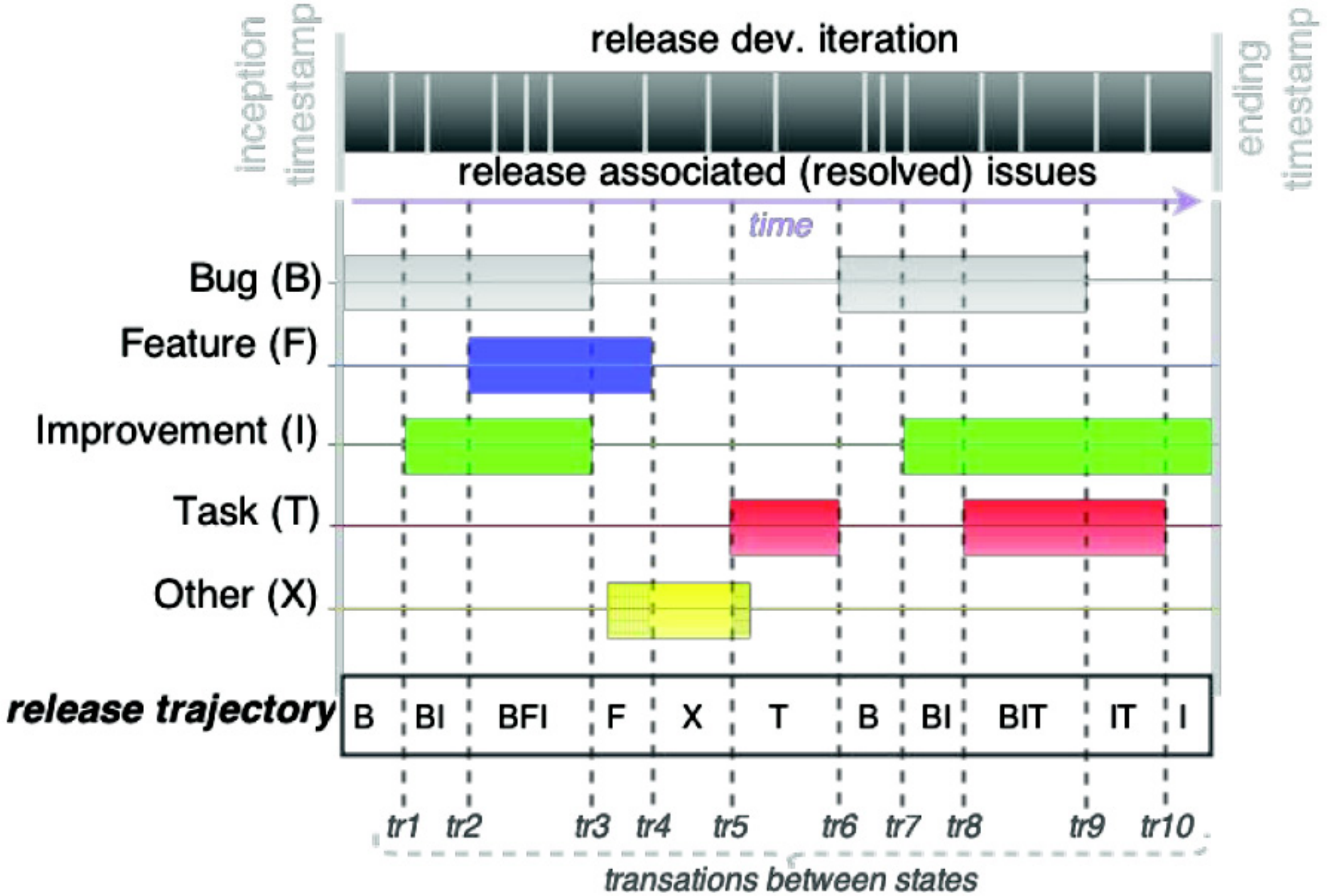}
\vspace{-0.3cm}
\caption{Overview of release trajectory modeling approach: from issues tracking processes to a trajectory of states.}
\label{fig:TrajectoryModelling}
\end{figure}

\Figref{fig:TrajectoryModelling} depicts our approach for mapping the closed resolved issues associated to a given release to a trajectory (sequence) of distinct states representing the release's trajectory. In this representation, each state has its specific duration, and the duration of the release's trajectory is the period between its inception date and ending date.
The modeling algorithm is summarized as follows:
\begin{itemize}[leftmargin=*]
	\item	
		\emph{Identify \textbf{atomic states and transitions} w.r.t. each considered issue type:}
		For each issue type, $p$, issues whose lifecycles overlap, regarding their creation and closing dates, are grouped together and aggregated to form one atomic state characterized by $p$.  That state is starting at an {\it opening timestamp} ($open\_t$), the earliest creation date of issues forming the state, and ending at a {\it closing timestamp} ($close\_t$), the latest closing date of issues forming the state.
A state can indeed be formed of one issue if it does not overlap with other issues of its type. In such a case, the state's $open\_t$ and $close\_t$ are respectively the creation date and closing date of that issue.
		These timestamps $open\_t$ and $close\_t$ represents two atomic transitions, that are respectively moving into (\ie opening) and moving out (\ie closing) the underlined state.
		The output of this step is a map between considered issue types and their lists of opening and closing transitions.
		In \Figref{fig:TrajectoryModelling}, the Bug issues which are associated to the analyzed release form two atomic Bug states. The first one goes from the inception timestamp to $tr3$ and the second starts at $tr6$ and ends at $tr9$. These states indicate periods when the undertaken release is diagnosed as suffering from bugs, and the period between transitions $tr3$ and $tr6$ shows that no bug issue associated to the release was open.
		Hence, from the perspective of the analyzed release, this period indicates a healthy/clean state, in which  the release developers did not diagnose bugs that may concern the undertaken release.
		
	\item	
		\emph{From \textbf{atomic} states and transitions to \textbf{global} ones w.r.t. all considered issue types:}
		In this step, the lists of atomic opening and closing transitions for all considered issue types are merged into one list, and sorted from the old one to the new one.
		Then, the algorithm processes the transitions so that atomic states which occur between two successive transitions from one global, aggregated, state delimited by those transitions.
		The output is then a list of ordered global states along with the transitions between those states (that is the union of all atomic transitions) forming the effective trajectory of the underlined release.
		By definition, a global state is said {\it atomic} if it concerns only of issue type, and is said {\it complex} if it concerns different types of issues.
		Consequently, the complexity of a state can be defined by the number of issue types that are involved in the state.
		In \Figref{fig:TrajectoryModelling}, the analyzed release is stated as suffering from bugs from its inception timestamp.
		At the transition $tr1$, the release is still diagnosed as suffering from bugs, and also certain improvements are required. Hence, $tr1$ is a transition from an atomic state, B, to a (more) complex one, BI = B + I.
		At $tr2$, another symptom is identified when the release requires additional features, whilst certain bugs fixing and improvements are still required. The release keeps this last state for a relatively long time until $tr3$ when both identified bugs and required improvements are sorted out, but the release still requires additional features. We observe that by the end of the release development iteration only improvements were required. 

	\item	
		\emph{Refine the release trajectory with regard to non-relevant issue types:}
		As mentioned earlier, issue types are sorted into two categories relevant/recurrent and non-relevant/non-recurrent. Issues which belong to non-relevant types are then then processed all together assigned to one virtual type, annotated by X in \Figref{fig:TrajectoryModelling}.
		Moreover,when an atomic state of the non-relevant type X overlaps with other states that are of relevant types, then the global state in the overlapping period is formed only by the states of relevant types. Hence, atomic states of the non-relevant type can impose global states only when other states do not occur.
		This aims at reducing noises of non-relevant issues on the release trajectory, as well as reducing the complexity of visualization techniques that can be used to analyze release trajectories.
		\Figref{fig:TrajectoryModelling} shows that there is one atomic state of the non-relevant type X, and this state overlaps with an atomic state of type F, between $tr3$ and $tr4$, and another atomic state of type T, between $tr5$ and $tr6$. However, the global states in those periods of overlapping are formed only by the relevant issue types F and T.
		The period between $tr4$ and $tr5$ is the only one with the global state of type X as no other relevant states occurred during that period. 	
		
	\item\emph{Normalize the duration of releases' trajectories:}
		A final, optional, step is to normalize the duration of releases' trajectories so that the duration of a state within a release trajectory is proportional to the release duration.
		Since releases can have completely different durations, this normalization is important when investigating recurrent patterns of developmental paths.
		In this paper, we normalize trajectories before analyzing them.
\end{itemize}



\begin{figure}[!t]
\centering
	\includegraphics[width=0.6\linewidth]{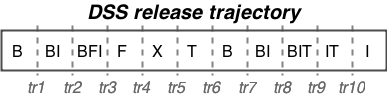}
\vspace{-0.4cm}
\caption{The DSS representation of the release's trajectory in \Figref{fig:TrajectoryModelling}.}
\label{fig:TrajectoryModelling_DSS}
\end{figure}

A release trajectory can also be summarized as a sequence of ordered distinct states where the states' duration is ignored.
This representation, namely \emph{\textbf{d}istinct-\textbf{s}uccessive-\textbf{s}tate} representation (DSS for short), is useful to focus on the number of states and transitions that occur in a release trajectory.
\Figref{fig:TrajectoryModelling_DSS} shows the DSS representation of the release's trajectory represented in \Figref{fig:TrajectoryModelling}, where the trajectory length reflects only the number of transitions in the release's trajectory.

\begin{figure}[!t]
\centering
	\includegraphics[width=0.65\linewidth]{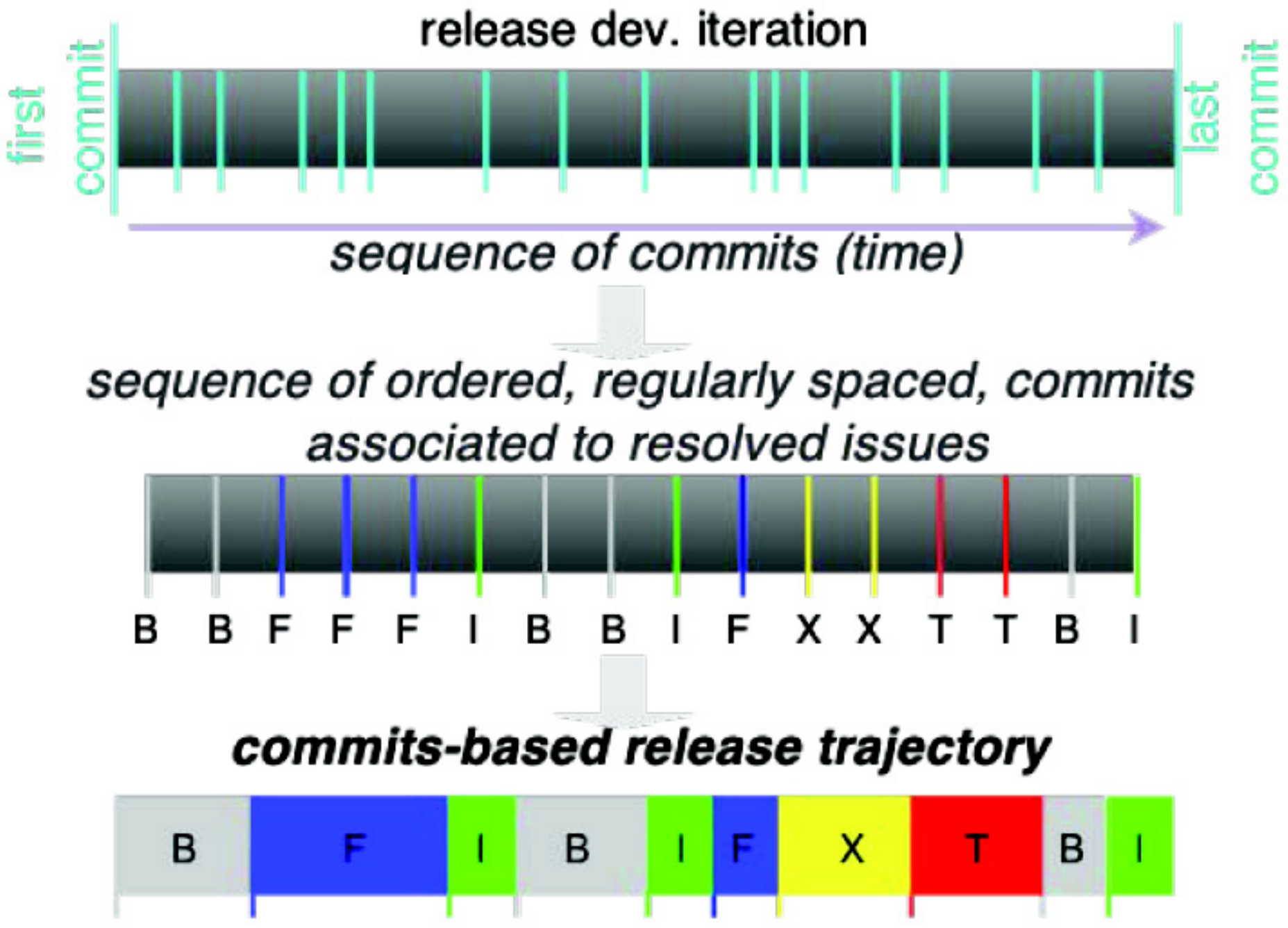}
\vspace{-0.3cm}
\caption{Example of commits-based release trajectory.}
\label{fig:TrajectoryModelling_Commits}
\end{figure}

\subsection{Commits-based Release Trajectory}\label{sec:commitsTrajectory}
Up to now, the representation of a release's trajectory is fully based on reported statuses of the release's associated issues in the issue tracking system.
This representation focuses on the temporal evolution of software states regardless of chronology of detailed changes that lead to changing the software states.
These detailed changes are recorded in historical code repositories, such as SVN and Git, in the project's change log which consists of a list of ordered commits.
Assuming that each commit in the project's log includes only changes for one specific task (\ie the project's commits do not include tangled changes \cite{Kiri14, Herz13}), and the project's recorded issues are tagged explicitly in the commits' logs, then both commits and types of recorded and tagged issues can provide further insight on releases' trajectory.
Consequently, a new kind of releases' trajectories can be constructed, namely {\it comments-based} (or {\it activity-based}) trajectories.

\Figref{fig:TrajectoryModelling_Commits} shows an example of commits-based release trajectory, where the release's development iteration is represented by a sequence of 17 commits, ordered chronologically. All of these commits are explicitly tagged in their logs by issues associated to the underlined release. Commits which cannot be connected to the release's issues are not considered.
In the first step, commits are annotated by the representative types of their associated issues.
Then, they are regularly spaced regardless of the elapsed time between them.
In other words, the unit of duration of commits-based releases' trajectories is ``commit'', not time.
For instance, the duration of the release's trajectory in \Figref{fig:TrajectoryModelling_Commits} is 17 commits.
The first two commits are annotated by the issue type Bug (B), the next three commits are annotated by the issue type new Feature (F), and so on and so forth until the last commit, annotated by the issue type Improvement (I).
Finally, successive commits which have the same annotation type, such as the first two commits (B), are merged to form one state, and the duration of the state is measured by the number of merged commits.
This leads to the commits-based release's trajectory which is represented similarly to the normal (issues-based) release's trajectory. The main differences in commits-based releases trajectories is that 1) the trajectory duration is measured by the number of involved commits, and 2) the trajectory's states are all {\it atomic} as states cannot overlap.

Similarly to normal releases' trajectories, the commits-based trajectories can also be normalized to have the same duration, so that the duration of a state within a commits-based trajectory is proportional to the number of commits involved in the trajectory.
Furthermore, commits-based trajectories can also have the DSS representation where the duration of states is ignored, and then the trajectory's length reflects the number of transitions between the trajectory's states.

\begin{table*}[!t]
	\caption{Information about collected data from study software projects}
	\vspace{-0.5cm}
	\label{table:projects}
	\newcolumntype{Z}{>{\centering\arraybackslash} m{2.2cm}}
	\newcolumntype{X}{>{\centering\arraybackslash} m{2.5cm}}
	\newcolumntype{Y}{>{\centering\arraybackslash} m{1.2cm}}
	\begin{center}
	\setlength\tabcolsep{2pt}
	\begin{tabular}{lZYZZYYXX}
		\toprule
		Project &
		{Releases}	&
		{\#Releases}   &
		{Period}   &
		{Avg \#days}\break per release&
		{\#Issues}\break -resolved-&
		{\#Commits}   &
		$\%$Issues\break tracked in\break Commits&
		$\%$Commits\break tagged by\break Issues
		\\
		\midrule
		Solr &
		1.1 $->$ 4.10.3   &
		32   &
		Jan-06 to Dec-14   &
		101 days   &
		3619   &
		9296   &
		$3036/3619 = 84\%$   &
		$4841/9296 = 52\%$
		\\
		\midrule
		OpenJPA &
		0.9.6 $->$ 2.3.0   &
		17	&
		Fev-06 to Nov-13   &
		161 days   &
		1553   &
		7219   &
		$1348/1553 = 87\%$	&
		$4042/7219 = 56\%$
		\\
		\midrule
		Struts2 &
		2.1.2 $->$ 2.3.21   &
		35	&
		Oct-07 to Dec-14   &
		72 days   &
		1317   &
		4325   &
		$916/1317 = 70\%$	&
		$1263/4325 = 29\%$
		\\
		\midrule
				\rowcolor{lightGray}
				&
				&
				84   &
				$\approx 7 years$   &
				&
				6489   &
				20840   &
				&
				
				\\
		\bottomrule
	\end{tabular}
	\end{center}
\end{table*}

\subsection{Summarizing Releases' Trajectories}\label{sec:trajectoryProperties}
Once releases' trajectories are constructed, sequence-analysis techniques can be applied for summarizing and analyzing the trajectories.
In this section, we describe the main techniques and properties that we believe relevant to the comprehension of software releases' trajectories.

First of all, each release trajectory is characterized by three basic properties:
1) distinct states involved in it (the states in the DSS format),
2) the number of transitions between those states,
and 3) the durations of those states.

Sequence-analysis provide a wide range of techniques for synthesizing global information for a family of trajectories (\eg the trajectories of releases of a considered software project), and computing overall descriptive statistics.
In our context, we mainly focus on techniques that allow software managers to understand the overall picture of releases' trajectories and capturing valuable pieces of information that can assist them while planning next releases.
More specifically, we focus on techniques for estimating the probability to switch at a given position from one state $s_i$ to another one $s_j$ (\ie transition rates between states), and for summarizing the trajectories of software releases in one picture, said the trajectory made
of the most frequent state at each position (\ie the trajectory of modal states).
\begin{itemize}
	\item{\it Transition rates: }
	The transition rate between two distinct states $s_i$ and $s_j$ is the probability to switch at a given position from $s_i$ to $s_j$, $p(s_j \mid s_i)$. This probability, as implemented in TraMineR \cite{Traminer}, is based on the occurrences of this transition in the analyzed trajectories as well as on the occurrences $s_i$ not followed by $s_j$.
It is important to note that the transition rates between two states (from $s_i$ to $s_j$, and from $s_j$ to $s_i$) are usually not equal.
	
	\item{\it The trajectory of modal states:}
		An interesting summary over the releases' trajectories is the trajectory made
of the most frequent state at each position $p$, with regard to all analyzed releases of a given software project.
		In the trajectory of modal states, each position is characterized by one state (\ie the most frequent state at that position) and the frequency of that state at that position.
		This trajectory provides an overall picture of the analyzed releases' trajectories, and shows at which positions the trajectories of analyzed releases behave similarly (the frequency of modal states at those positions is high), and at which positions they behave differently (where the frequency of modal states is relatively low).   	
\end{itemize}

\vspace{+1.2cm}
\subsection{Mining Patterns of Releases' Trajectories}\label{sec:patternsMining}

Although the modal --most frequent-- trajectory may provide a useful summary over a family of releases' trajectories, it is still of limited interest since it captures only one general pattern of the family of the analyzed releases.
A more useful approach consists in identifying sets of distinct (sub) families of trajectories and identifying recurrent patterns of releases' trajectories.
For this purpose, clustering methods can be employed to identify homogeneous clusters of releases' trajectories.
To this end, we need techniques that can assess similarities/distances between releases' trajectories.

Several sequence similarity (dissimilarity) measures have been proposed in the literature.
Those measures can be classified into two categories:
1) based on the count of matching states and sub-sequences between the two compared trajectories, and 2) based on the minimal cost of transforming one of the compared trajectories into the other.
One of the most used dissimilarity measure for identifying groups of life-course trajectories with similar patterns is the Optimal Matching (OM) metric, that belongs to the second category.
For computing the OM between two trajectories, the minimal cost for transforming one into the other should be assessed. This cost depends on the allowed transforming operations and their individual costs.
The OM distance metric allows two transforming operations:
1) the substitution of one state by another one, and
2) the insertion or deletion of a state.
Hence, the OM between two trajectories $Tra_1$ and $Tra_2$ is defined as the minimal cost (in terms of insertions, deletions and substitutions) for transforming $Tra_1$ into $Tra_2$ or $Tra_2$ into $Tra_1$.

Usually, the cost of insertion or deletion operation of any state is set as a constant, independently of the state and its position in the trajectory.
As for the costs of substitution operations are depending on the concerned states, and hence they are usually organized in a square symmetrical matrix of dimension $A$, namely the substitution cost matrix (SCM), where $A$ is the size of the alphabet of possible distinct states in the analyzed trajectories.
In our context, for issues-based releases' trajectories, $A = 16$.
SCM can be specified manually, which can be effort-prone and error-prone activity.
Hence, it is recommended to generate the SCM automatically using statistical descriptive information about the relations between the distinct states in the alphabet.
In this paper, the substitution costs for two states $s_i$ and $s_j$ are determined from the estimated transition rates as $2 - p(s_i \mid s_j) - p(s_j \mid s_i)$, the cost is high when changes between $s_i$ and $s_j$ are rarely observed and lower when they are frequently observed.

\section{Evaluation}\label{sec:evaluation}
Using sequence-analysis of releases' trajectories, one can address important questions, such as:
``do releases' trajectories of a given software project follow the same development standard or methodology?",
``what are the frequent typologies of releases' trajectories that can represent standards?",
``when and why some releases follow turbulent chaotic trajectories or persist in one fixed state?"
and finally, ``what are the relations between the topological and longitudinal properties of releases' trajectories and other measurable properties of releases?'' 

To address these questions empirically, a large set of release trajectories must be analyzed.
Here, the contribution of sequence-analysis techniques is to assist in performing exploratory tasks that consist of summarizing the dataset, extracting relevant information to the question of interest, and organizing the trajectories into representative homogeneous groups that capture patterns of trajectories.
Then, the resulting summaries and patterns can be used in classical statistical analysis techniques, such as in regression-like models.


In this section, we evaluate the utility of our approach via a descriptive study of releases' trajectories using a fairly large dataset of releases, extracted from the historical evolution of three large software projects. Our study aims to put in evidence the utility of our approach for understanding the developmental paths of releases and extracting new kinds of reports that can address some of the aforementioned questions.

\vspace{-0.2cm}
\subsection{Projects and Data Collection}\label{sec:evaluation_Projects}
\vspace{-0.1cm}
We extracted the data from three large Apache projects, namely Solr \cite{Solr}, OpenJPA \cite{OpenJPA} and Struts \cite{Struts2}.
The issues of analyzed projects are collected from JIRA (http://issues.apache.org/jira/), and the change logs for Solr and OpenJPA are collected from their SVN repositories, while the change logs for Struts2 are collected from its Git repository.
In total, we collected issue and commit data associated to 84 releases of those projects, covering an evolution period of around 7 years for each project.
\Tabref{table:projects} shows for each project
the number of considered releases,
the average of releases' duration in days,
and the number of issues and commits associated to those releases.
As it can be seen, a large percentage of the considered issues are explicitly referenced in the commits' logs of the analyzed projects.
However, the percentage of commits that are explicitly tagged with issues is relatively small.
For instance, only 52\% of the Solr analyzed commits are tagged with issues that can explicitly explain the nature of development activity associated to them. 
In this study, only these commits can be used in our approach for modeling the commits-based releases' trajectories.

\begin{figure*}[t!]
\centering
	\includegraphics[width=0.4\linewidth]{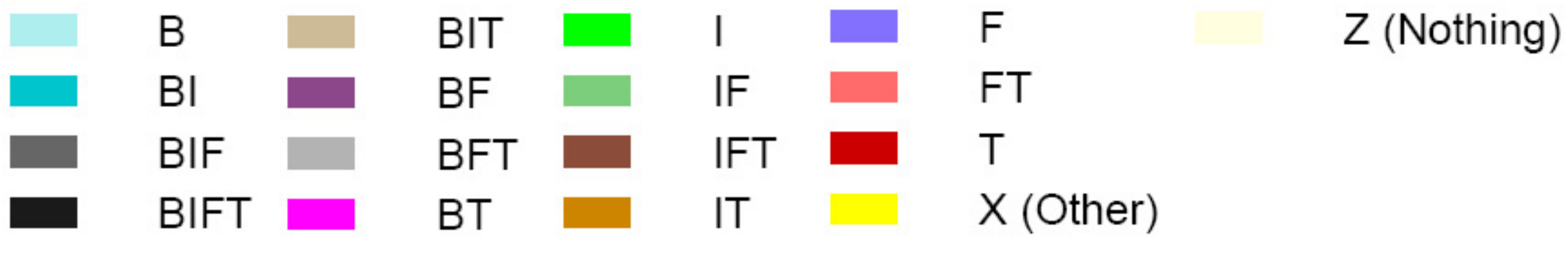}
	\\
	\vspace{-0.1cm}
	\subfigure[Solr]{
	\label{sfig:Solr_sequences}
	\includegraphics[width=0.25\linewidth]{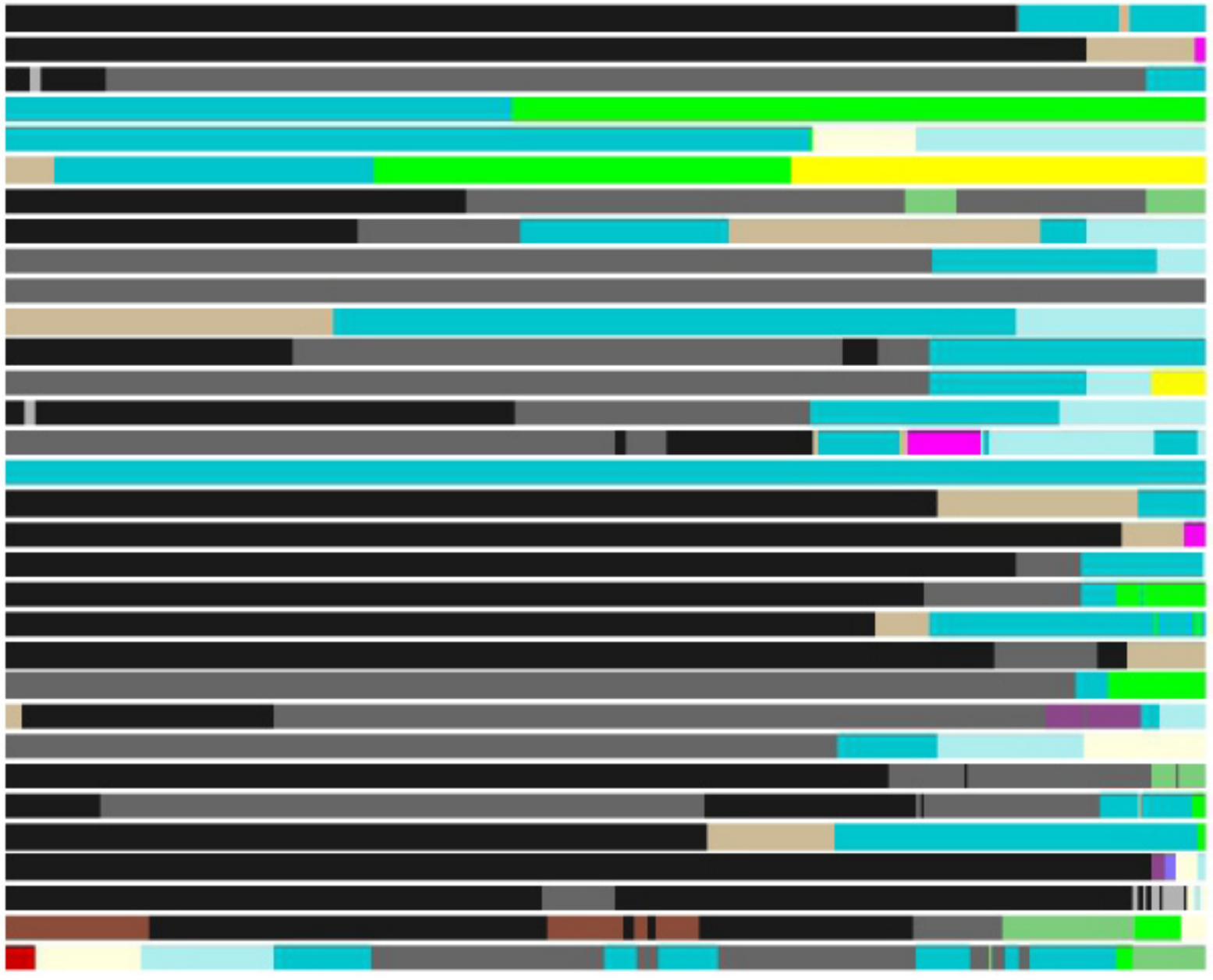}
	}
	\hfill
	{
		\subfigure[OpenJPA]{
		\label{sfig:OpenJPA_sequences}
		\includegraphics[width=0.25\linewidth]{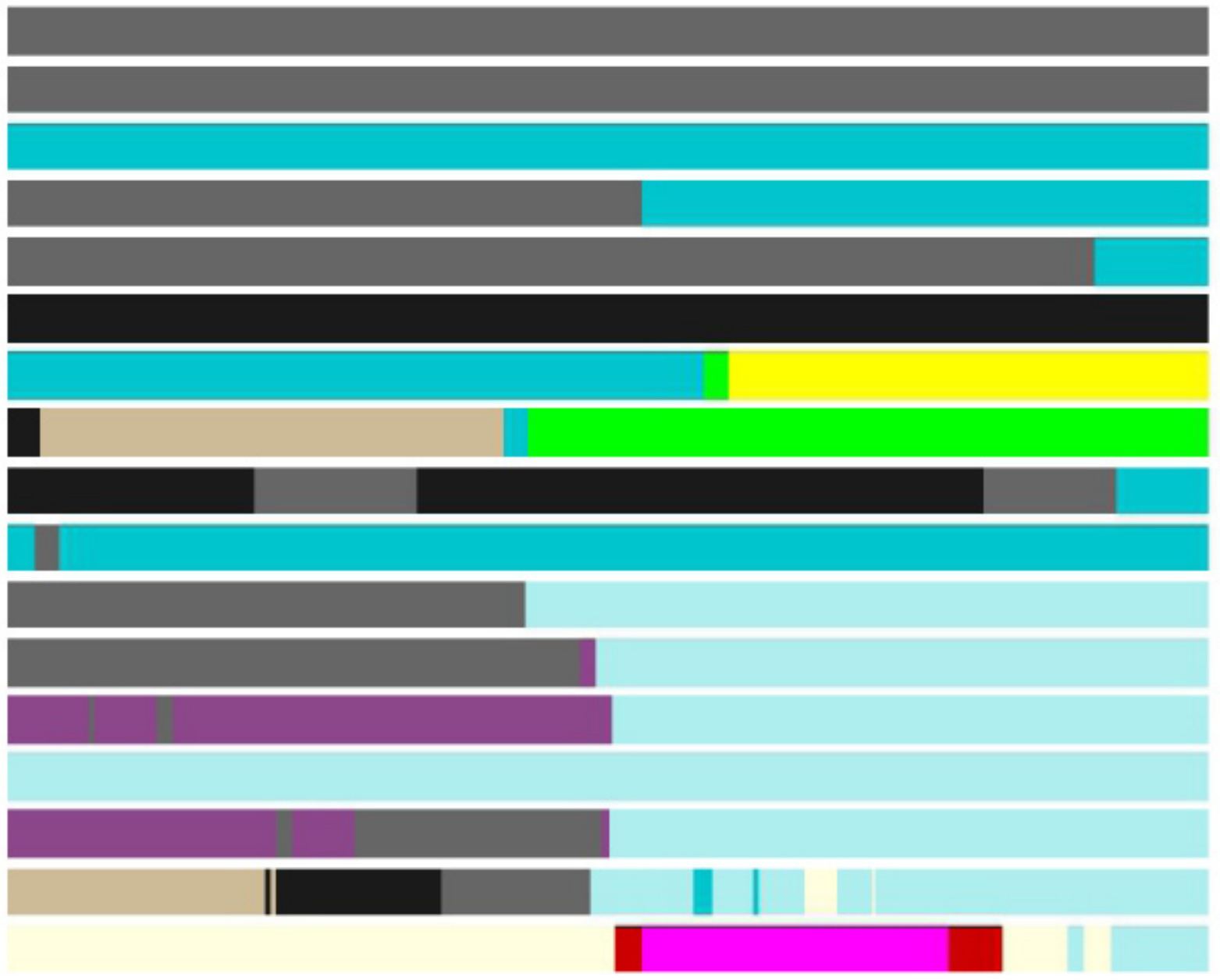}
		}
		\hfill
		\subfigure[Struts2]{
		\label{sfig:Struts2_sequences}
		\includegraphics[width=0.25\linewidth]{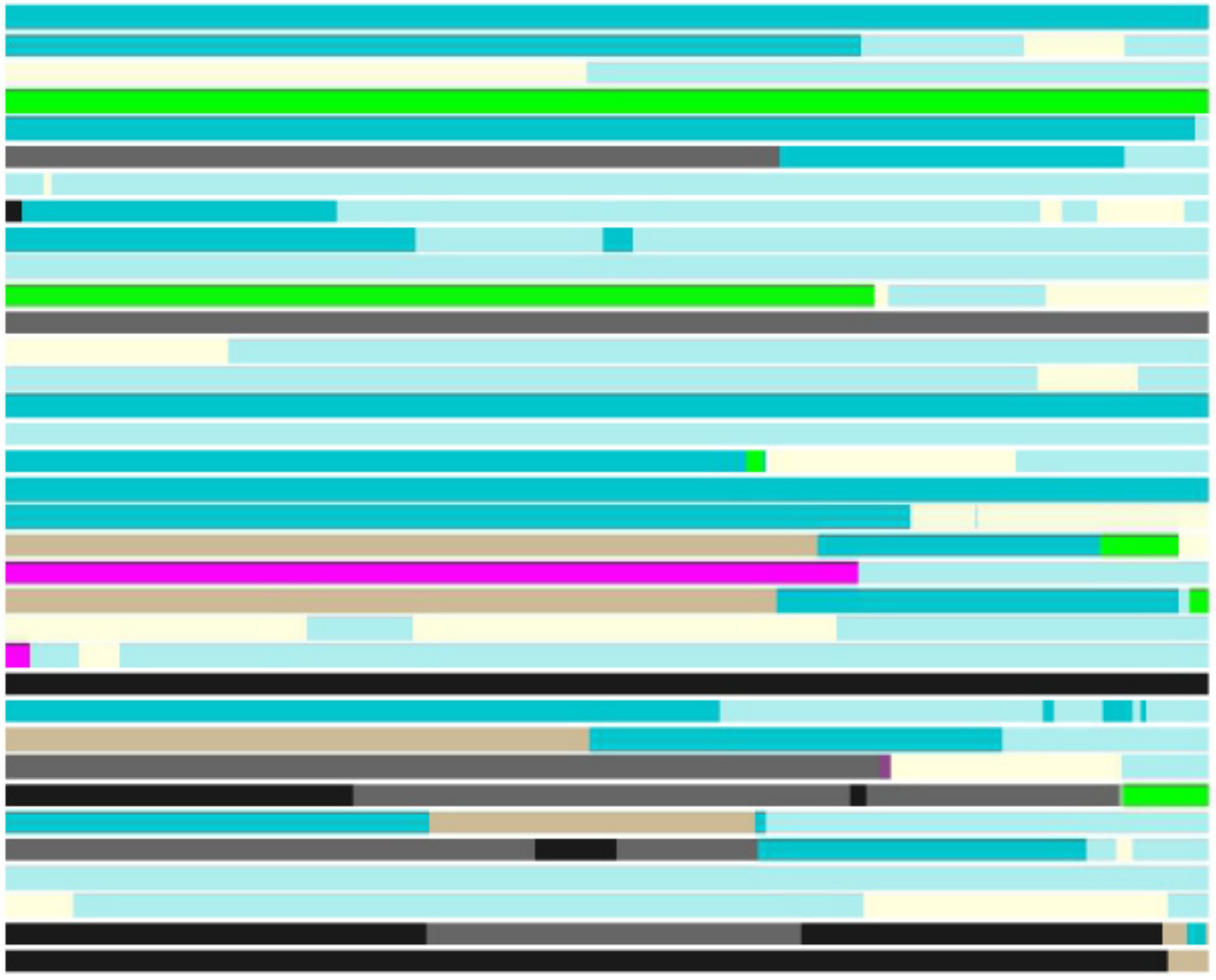}
		}
	}
\vspace{-0.3cm}
\caption{Trajectories of analyzed releases}
\label{fig:sequences}
\end{figure*}

\begin{figure*}[t!]
\centering
	\subfigure[Solr]{
	\label{sfig:Solr_comsequences}
	\includegraphics[width=0.25\linewidth]{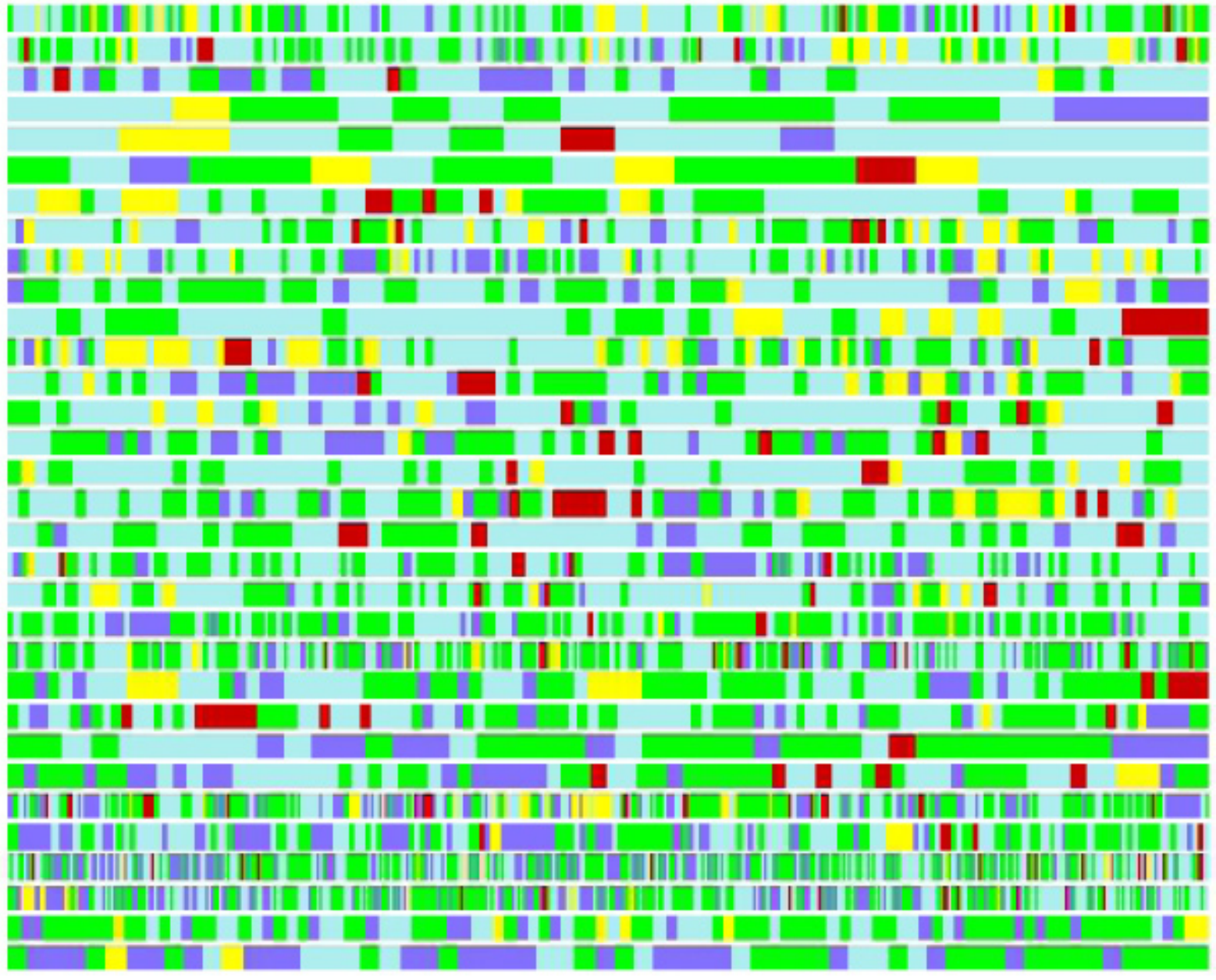}
	}
	\hfill
	{
		\subfigure[OpenJPA]{
		\label{sfig:OpenJPA_comsequences}
		\includegraphics[width=0.25\linewidth]{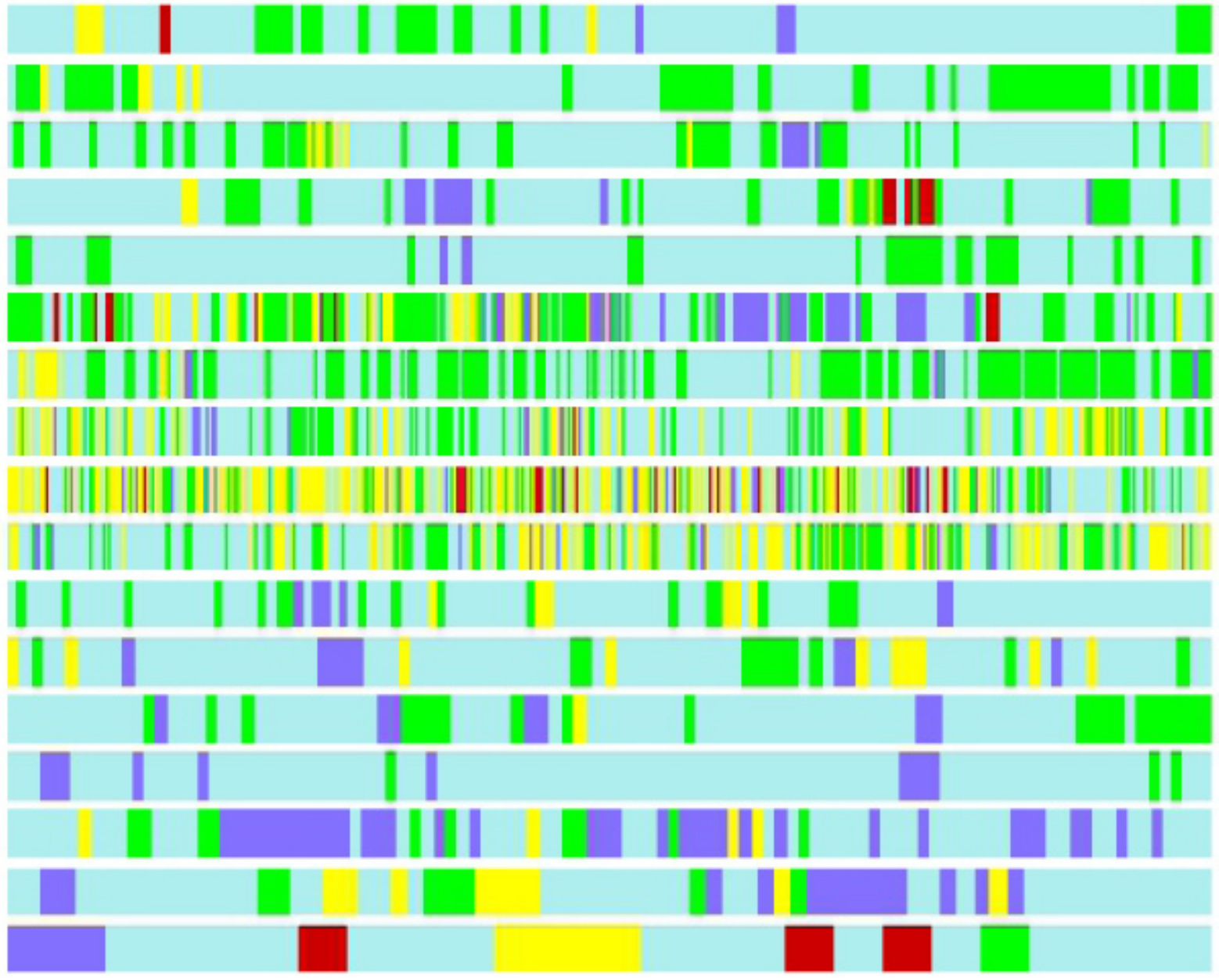}
		}
		\hfill
		\subfigure[Struts2]{
		\label{sfig:Struts2_comsequences}
		\includegraphics[width=0.25\linewidth]{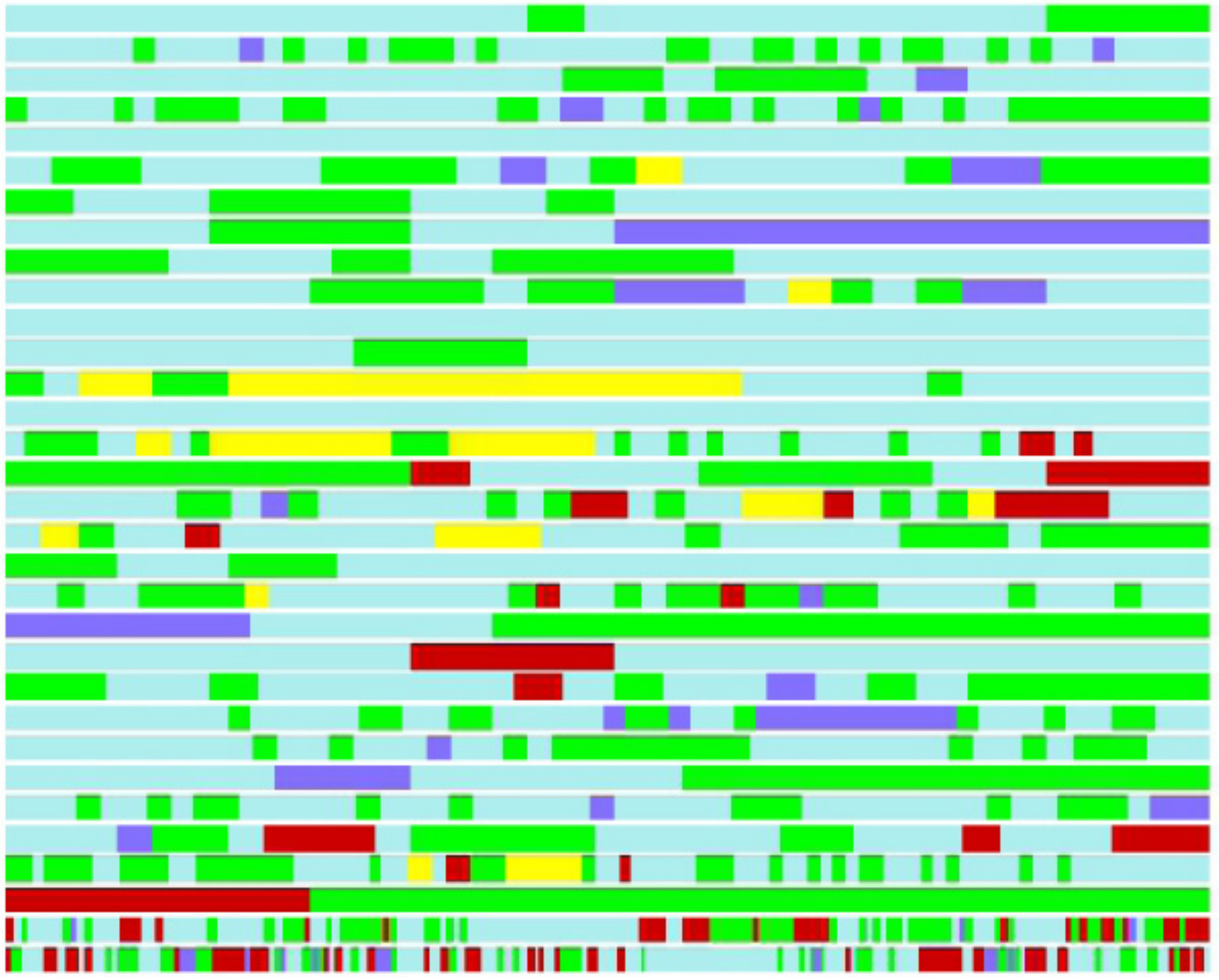}
		}
	}
\vspace{-0.3cm}
\caption{Commits-based trajectories of analyzed releases}
\label{fig:sequences_com}
\end{figure*}

\begin{figure*}[t!]
\centering
	\subfigure[Solr]{
	\label{sfig:Solr_sequences_dss_freq}
	\includegraphics[width=0.25\linewidth]{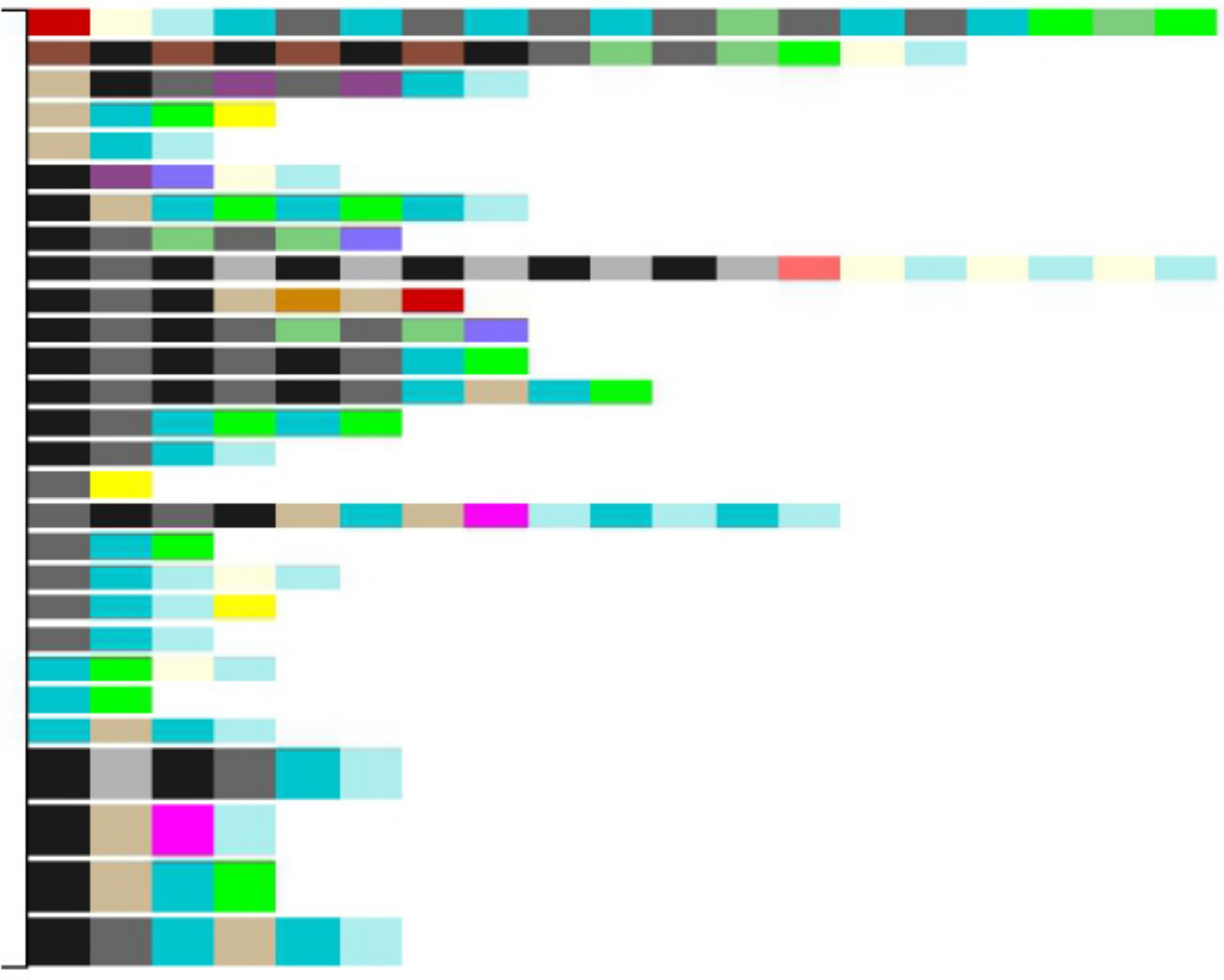}
	}
	\hfill
	{
		\subfigure[OpenJPA]{
		\label{sfig:OpenJPA_sequences_dss_freq}
		\includegraphics[width=0.25\linewidth]{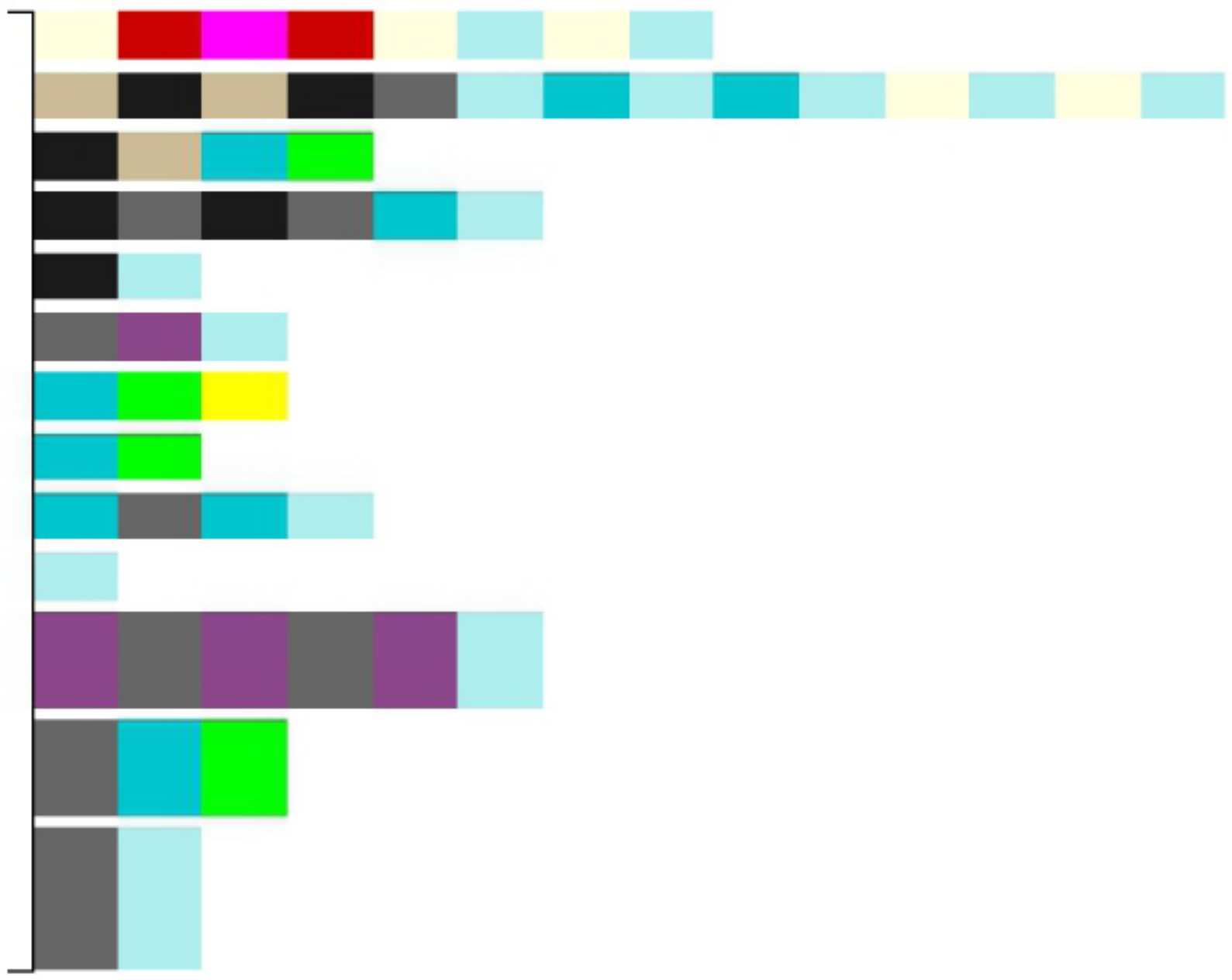}
		}
		\hfill
		\subfigure[Struts2]{
		\label{sfig:Struts2_sequences_dsss_freq}
		\includegraphics[width=0.25\linewidth]{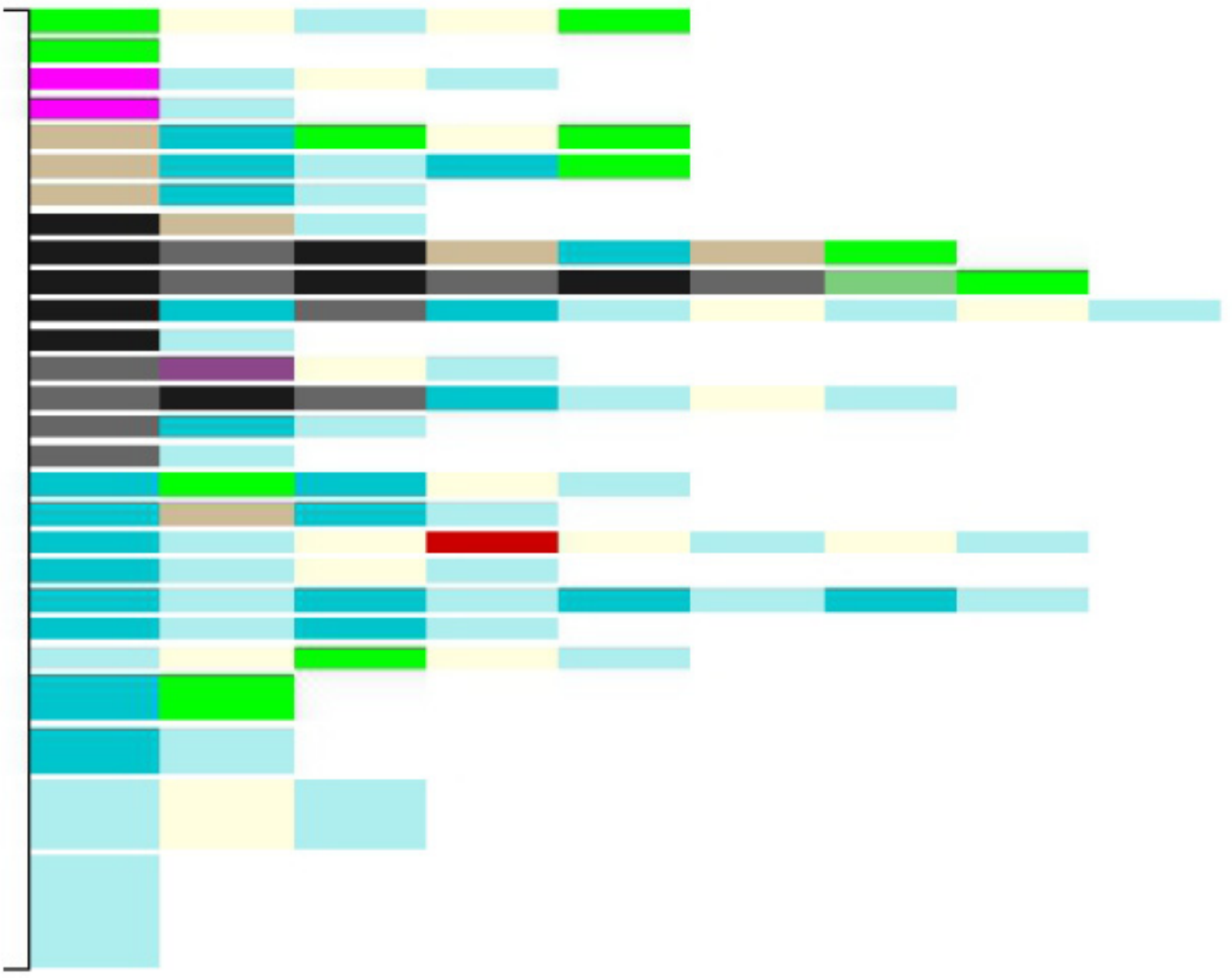}
		}
	}
\vspace{-0.3cm}
\caption{Cumulative frequency of of releases' trajectories in Distinct Successive State (DSS) format}
\label{fig:sequences_dss_freq}
\end{figure*}

\begin{figure*}[t!]
\centering
	\subfigure[Solr]{
	\label{sfig:Solr_sequences_dss_modal}
	\includegraphics[width=0.3\linewidth]{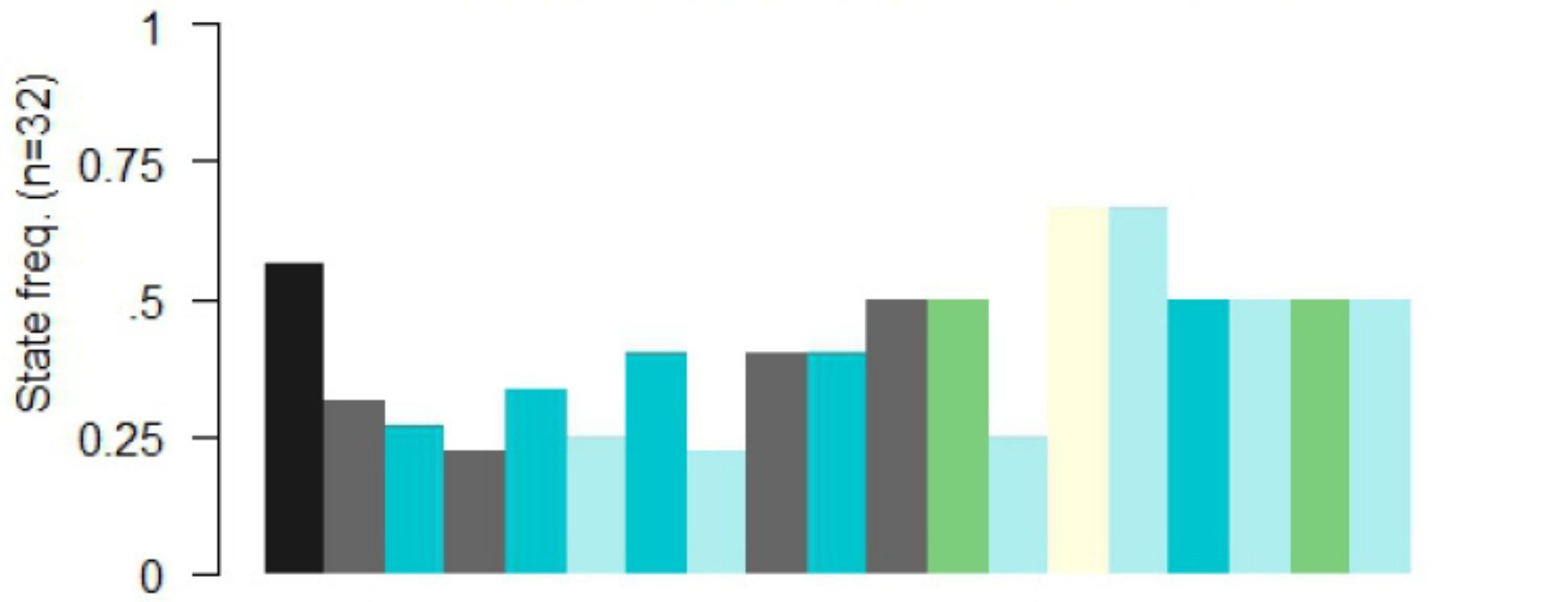}
	}
	\hfill
	{
		\subfigure[OpenJPA]{
		\label{sfig:OpenJPA_sequences_dss_modal}
		\includegraphics[width=0.3\linewidth]{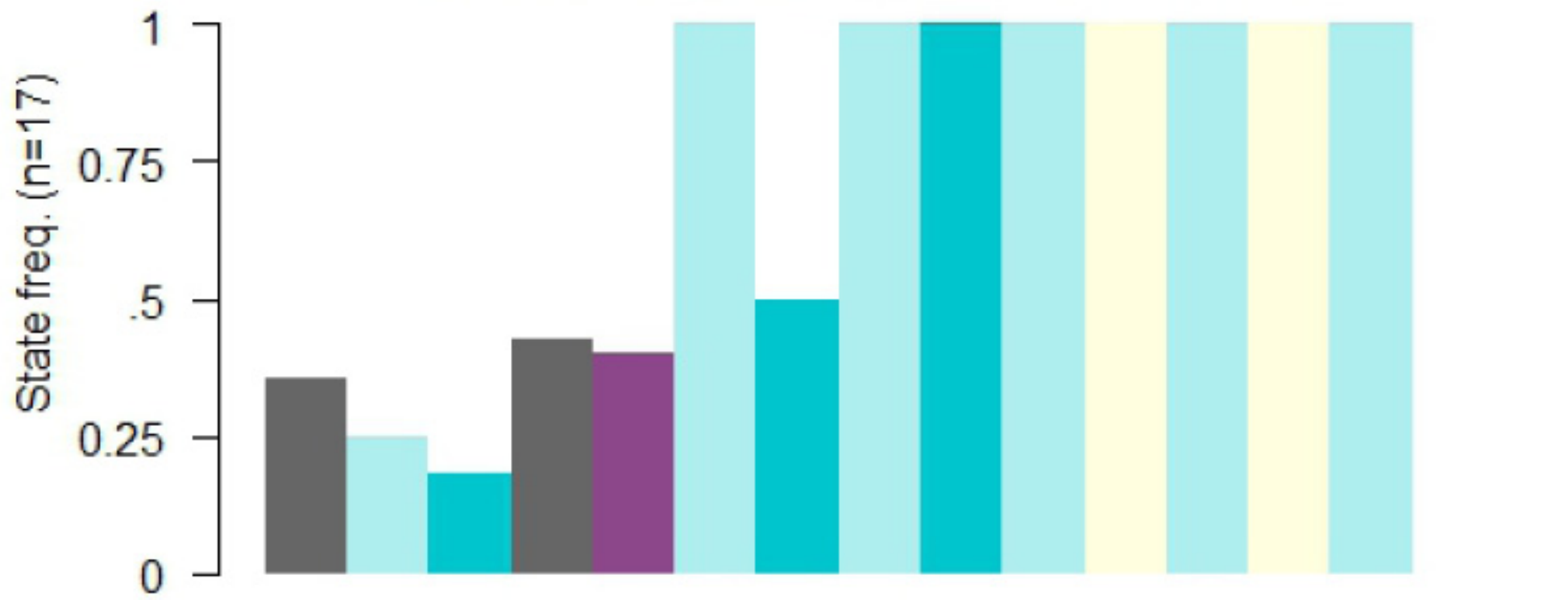}
		}
		\hfill
		\subfigure[Struts2]{
		\label{sfig:Struts2_sequences_dsss_modal}
		\includegraphics[width=0.3\linewidth]{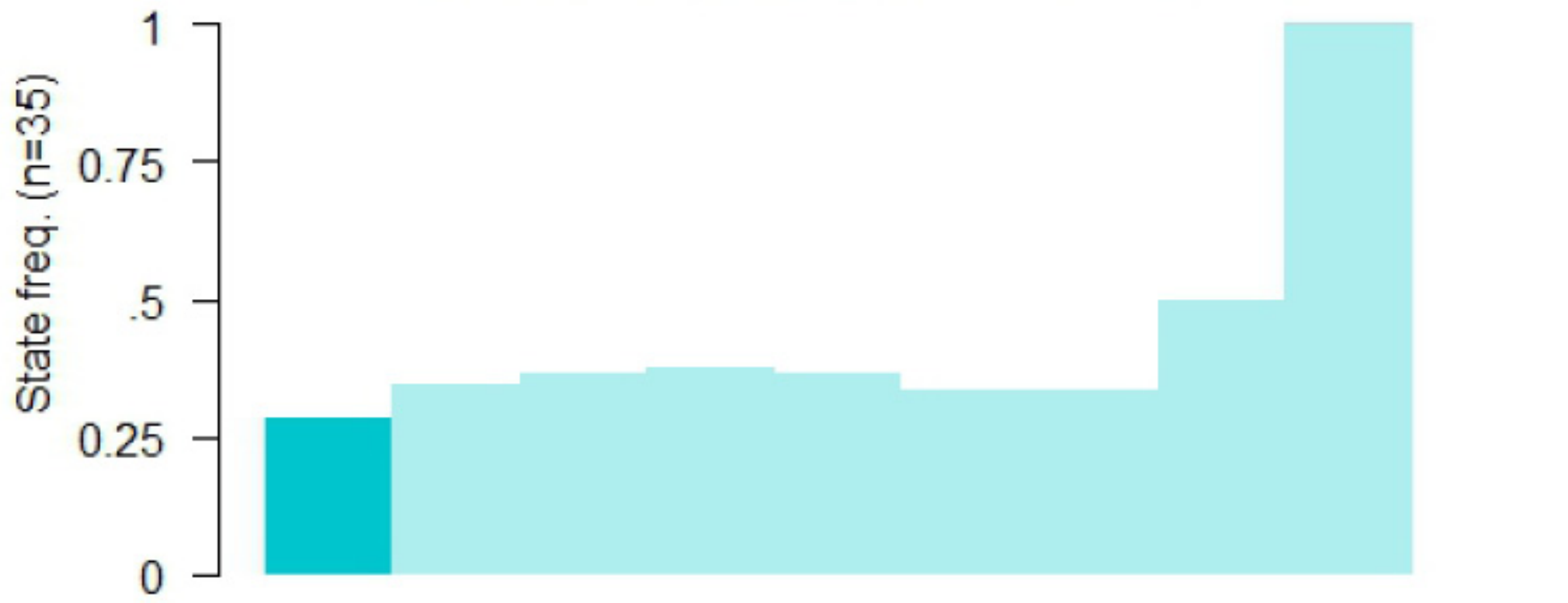}
		}
	}
\vspace{-0.3cm}
\caption{Modal --most frequent-- states' trajectory based on DSS format of releases trajectories}
\label{fig:sequences_dss_modal}
\end{figure*}

\begin{figure*}[t!]
\centering
	\subfigure[Solr]{
	\label{sfig:Solr_sequences_transitions}
	\includegraphics[scale=0.4]{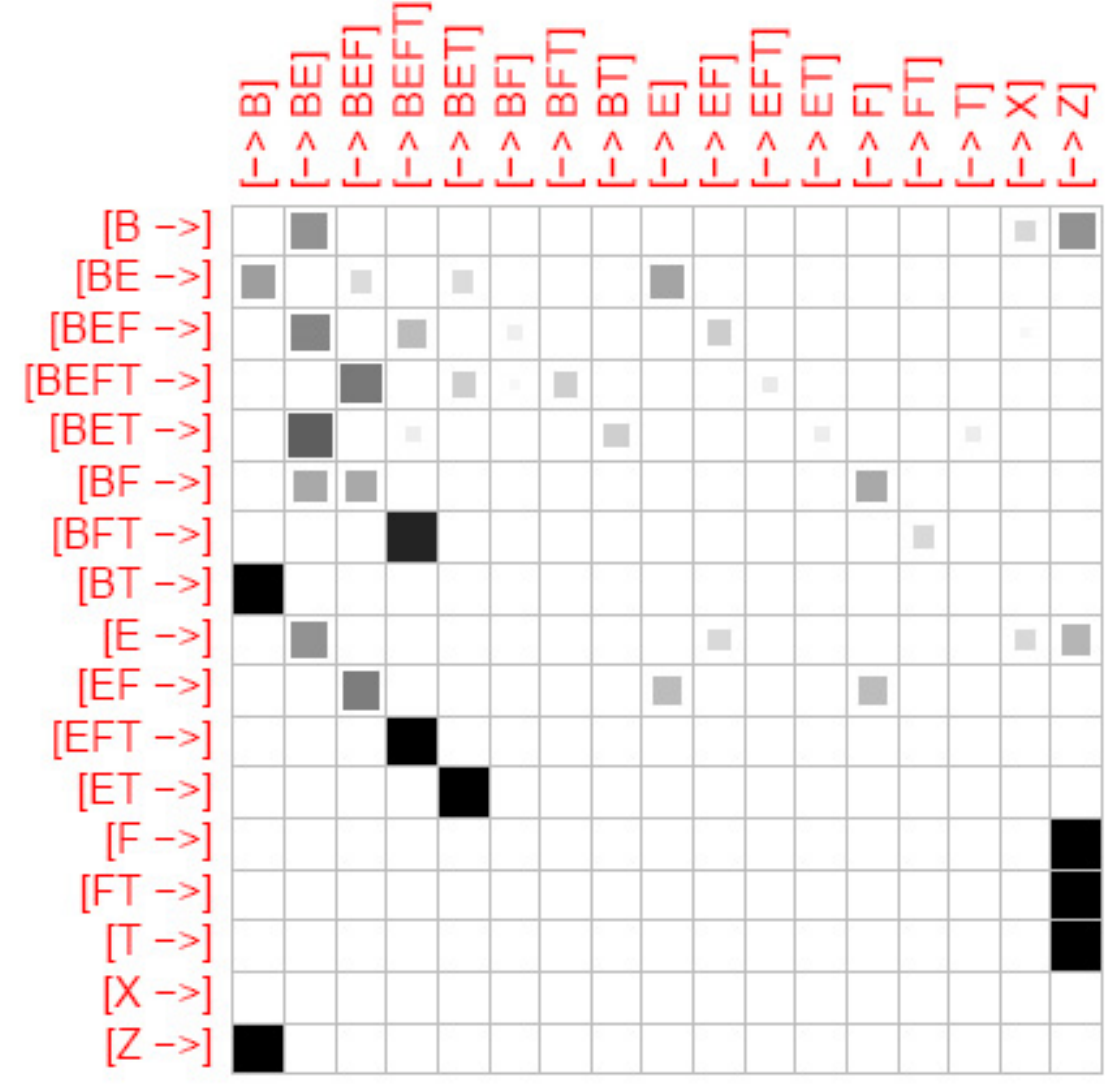}
	}
	\hfill
	{
		\subfigure[OpenJPA]{
		\label{sfig:OpenJPA_transitions}
		\includegraphics[scale=0.4]{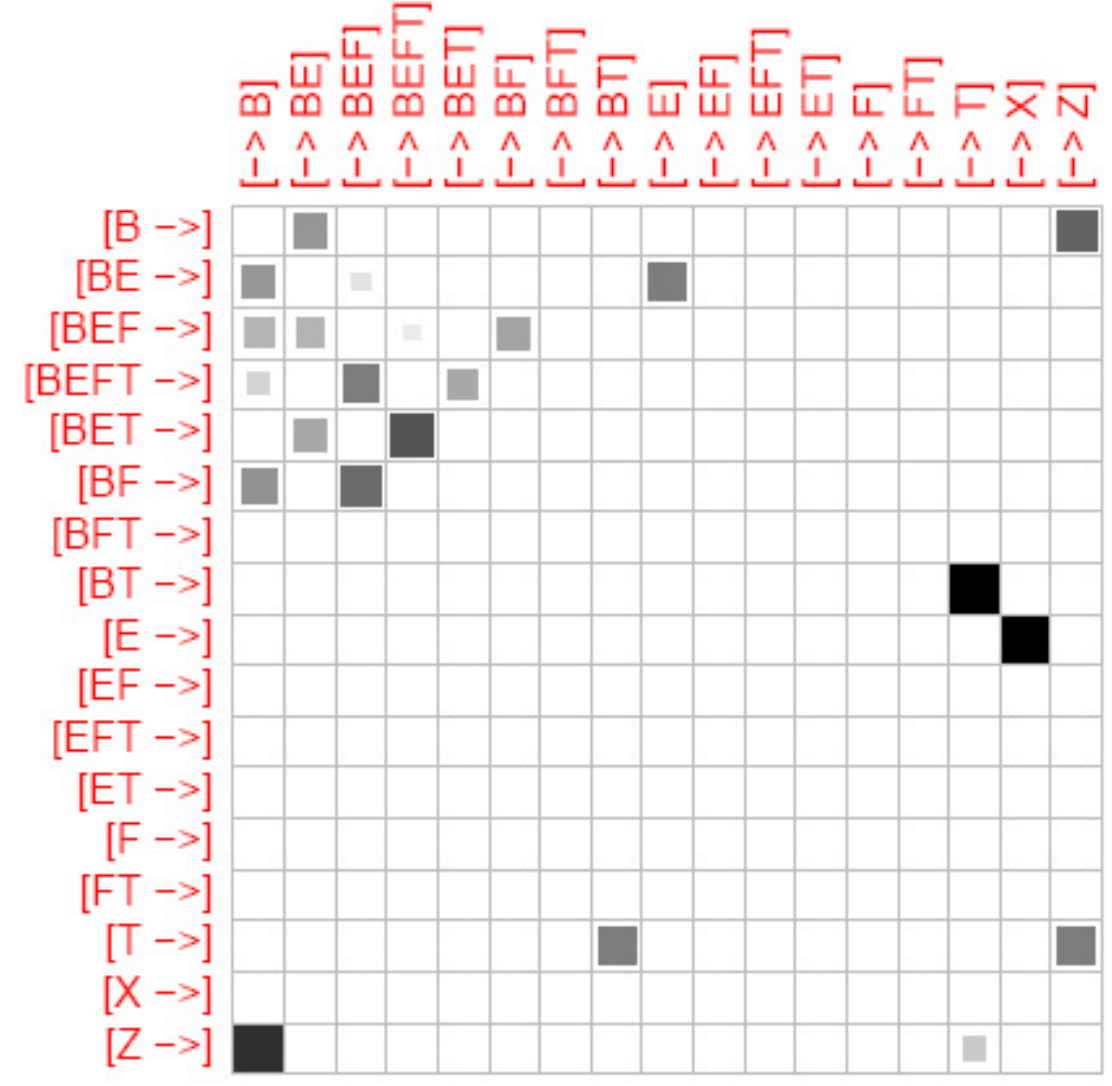}
		}
		\hfill
			{
			\subfigure[Struts2]{
			\label{sfig:Struts2_transitions}
			\includegraphics[scale=0.4]{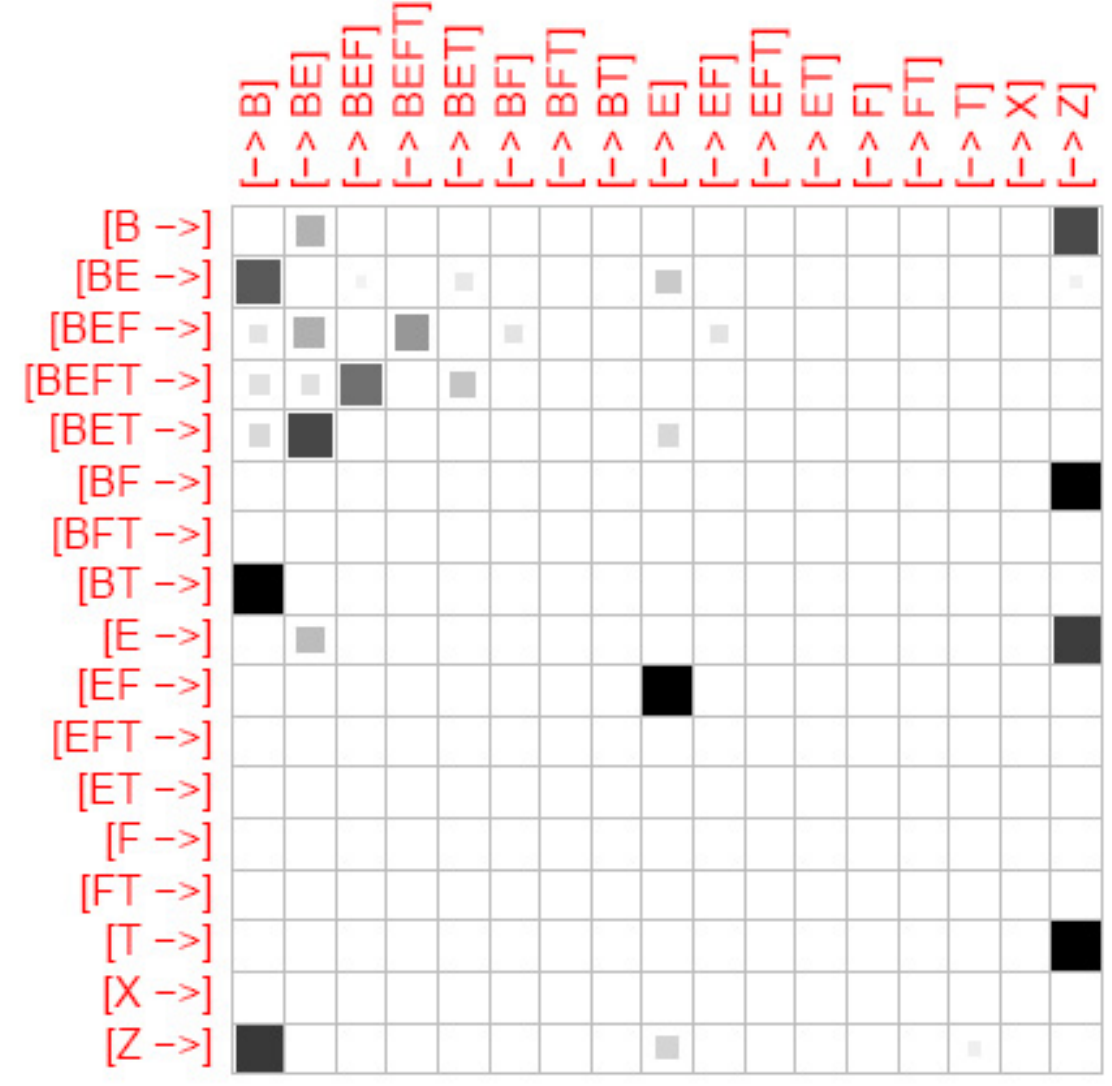}
			\hfill
			\includegraphics[scale=0.4]{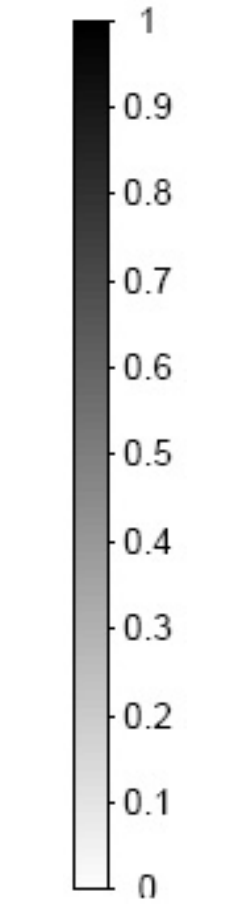}
			}
		}
	}
\vspace{-0.3cm}
\caption{Transition rates (probabilities) between distinct states in releases trajectories}
\label{fig:sequences_transitions}
\end{figure*}

\subsection{Overview of Releases' Trajectories}\label{sec:evaluation_OverviewTrajectories}
\Figref{fig:sequences} shows the trajectories of the analyzed releases.
In this figure, we easily observe that there are different shapes of releases' trajectories, particularly when comparing the releases of the three analyzed projects.
For instance, at a first glance on releases of the Solr project \Figref{sfig:Solr_sequences}, we observe that for almost all releases the complex states BIFT (black color state) and BIF (gray color state) are omnipresent along the trajectories.
This reflects that the Solr software is frequently in complex states whereas issues of all considered types (bug, B, improvement, I, new feature, F, and technical task, T) stay cohabiting for a long time during the development process, and developers work on these different issues in a parallel fashion.
More precisely, the commits-based releases' trajectories for Solr in \Figref{sfig:Solr_comsequences} shows that the development activities shift very frequently from one activity type to another one.
Here, we merely observe in some places a continuity in the nature of performed commits.
Indeed, only in some releases, we can observe that successive commits have focused on bug fixing (turquoise color state) or on performing some improvement (green color state).

Fortunately, the releases of the Struts2 project have different shapes than the ones of Solr, as shown in \Figref{sfig:Struts2_sequences} for the issues and \Figref{sfig:Struts2_comsequences} for the commits.
This reflects that the Solr and Struts2 projects adopt different strategies of development.
In Struts2, we observe that almost all releases are about fixing bugs and performing improvements, except for a small number of releases, especially the first two releases (at the bottom of figures). The trajectories of these two releases are of similar nature of the Solr releases.
In fact, these releases, $2.1.2$ and $2.1.3$, are among the earliest releases of the Struts Action Framework 2.0. They represent an important transition in the life cycle of Struts as they demarcate the period when Struts and WebWork projects have decided to join in one project, that is Struts2.
In the Struts2 project, we also observe that in some releases, technical task and/or new feature inquiries cohabit with bug fixing ones (see fuchsia and gray color states). However, we can observe in the commits-based trajectories that development activities associated to these inquiries are performed in a relatively continuous fashion.

As for the OpenJPA project, we observe in \Figref{sfig:OpenJPA_sequences} and \Figref{sfig:OpenJPA_comsequences} that the releases' trajectories can be classified into two categories. In the first category, the releases' trajectories are similar to those of the Solr project, whereas issues of all types cohabit together for long periods, and the development activities shift very frequently from one development nature to another.
These releases occur mainly after the mid life of OpenJPA, precisely, when OpenJPA has passed from $1.x.x$ to $2.x.x$. After this transition, a relatively large number of new features have been incorporated in OpenJPA (mainly in $2.1.0$) and a large number of improvements have been performed\footnote{Refer to http://openjpa.apache.org/builds/2.0.0/apache-openjpa-2.0.0/RELEASE-NOTES.html}.
In the second category, which involves the earliest releases and $1.x.x$ releases, the releases' trajectories are more similar to those of the Struts2 project. Still, the turbulence of development activity, commits-based, states is more visible in OpenJPA than in Struts2.

Regardless of the states' duration in the trajectories, \Figref{fig:sequences_dss_freq} shows the distinct successive state (DSS) format of releases' trajectories.
We easily observe that the projects Solr and Struts include both long and short trajectories, whilst almost all trajectories of OpenJPA are short.
We also observe that the releases of Struts2 project are different from those of Solr and OpenJPA as they (almost all) begin with relatively simple states, that are only requesting bug fixing and/or improvements.
Moreover, a distinguishable property of Struts2 releases' trajectories is that many releases
have periods where no issue is opened, indicated by the letter Z for Zen (or clean) state, and thereafter bug or improvement issues appear. These states may reflect periods of relaxation, or reflection, in releases' development.
The DSS trajectories, can be summarized by the modal --most frequent-- states trajectories as in \Figref{fig:sequences_dss_modal}.
Here, the Solr modal trajectory in \Figref{sfig:Solr_sequences_dss_modal} shows that more than 50\% of Solr releases begin with the complex state which is associated to all types of considered issues, the state BIFT. However, this state is shared only in the beginning of releases' trajectories, and most releases end with states related to bug issues.
We also observe that at each position, at least 25\% of the Solr releases share one, modal, state.
The same observations can be reported to the modal trajectory of OpenJPA, as shown in \Figref{sfig:OpenJPA_sequences_dss_modal}. Unlike both Solr and OpenJPA, \Figref{sfig:Struts2_sequences_dsss_modal} shows that the modal state in almost all positions in the Struts2 trajectory is related to bug issues only. Still that, the frequency of this modal state in not high in almost all the positions.


\begin{figure*}[t!]
\centering
	\includegraphics[width=0.70\linewidth]{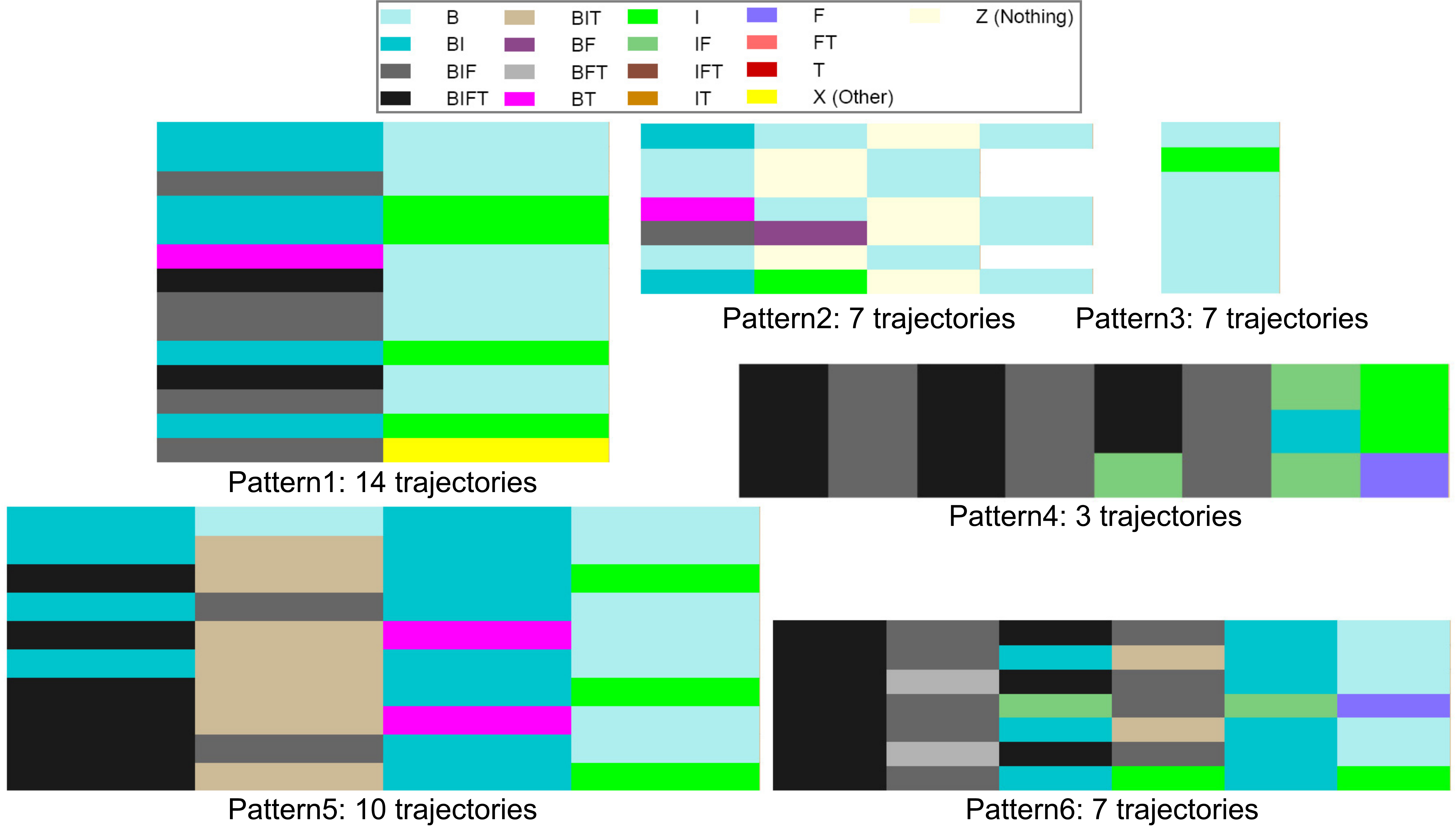}
\vspace{-0.1cm}
\caption{Identified patterns of releases' trajectories --based on DSS format and the transition rates between distinct states with regard to all analyzed trajectories--}
\label{fig:sequences_patterns}
\end{figure*}

\subsection{Transition Rates}\label{sec:evaluation_TransitionRates}
An important information that helps software managers in better understanding the evolution of software states through a new release is around the frequency and consistency of transitions between different states.
That is the transition rate, or probability, between software states.
\Figref{fig:sequences_transitions} shows in form of matrix the probabilities of transitions between different states in the releases' trajectories of the three analyzed projects.
The first important information that we observe is that the transition probability between the same two states can considerably differ from one project to another one.
This reflects that the transition probability between two states cannot be generalized with regard to the nature of states. It rather depends on the evolution path of the analyzed project and its development methodology.

In the Solr project, \Figref{sfig:Solr_sequences_transitions}, we observe several decisive transition probabilities showing that the involved states are strongly connected in the evolution process of software states.
For instance, we observe that when Solr state is BT (\ie requesting bug fixing and performing technical task) the probability is then very strong to sort out the technical tasks first (the software will move to the state B).
However, if new features are requested in addition to BT (\ie the Solr state is BFT), then the probability is very strong that Solr will move the most complex state requesting software improvement/enhancement (move to the BIFT state).
The transition to the BIFT state also seems very probable if Solr is in the state IFT.
Hence, we deduce that when the states F and T cohabit together with another state, either B or I, then Solr will move to the most complex state BIFT, whereas all issues of all kind cohabit together.
Finally, we observe that when the state of Solr is only around new feature issues (F) or technical tasks issues (T), then it is very probable to sort out these issues before opening new ones, and Solr will move to a clean state (Z). However, if Solr is in a clean state (Z), then it is very probable that the first issues to be opened are bug issues, and Solr will move the state B.

In the same vein, \Figref{sfig:OpenJPA_transitions} shows that when the release state is concerned about bug and technical issues (BT) the probability is then strong to close bug issues first (move to T state).
However, this is not the case in Struts2 as \Figref{sfig:OpenJPA_transitions} shows, where the probability is high to close technical issues before bug issues.
We also observe that when OpenJPA is concerned about only improvement/enhancement issues, then it is most probable to close these issues and open directly other, non-recurrent, types of issues (move to X state).


\subsection{Patterns of Releases Trajectories}\label{sec:evaluation_Patterns}

In this section, we attempt to identify recurrent paths, or sub-paths, of software state evolution that can help in categorizing releases' trajectories in a limited number of groups. Such groups reflect recurrent patterns of releases' trajectories.

To identify patterns of releases' trajectories, we performed an automatic analysis of the 84 trajectories of the three projects, using clustering techniques.
The aim is to identify families of homogeneous releases' trajectories that can represent patterns in the universe of software releases.
In our analysis, we use the Optimal Matching (OM) distance metric, that we presented in \secref{sec:patternsMining}.
Recall that OM between two trajectories relies on the cost of transformations that should be applied to transform one trajectory into the other.
To estimate this cost, we used the transition probabilities between distinct states based on all analyzed releases' trajectories.
In this way, trajectories which are grouped (clustered) together are not necessarily supposed to be {\it identical} or to share {\it identical} relatively long subsequences of states.
Indeed, this criterion is relaxed so that two {\it distinct} states can be considered very similar if the transition probability between them is high, in the analyzed population of releases' trajectories.

\Figref{fig:sequences_patterns} summarizes the results of our analysis. Six groups of homogeneous releases' trajectories were identified, involving in total 48 releases.
These groups can be placed in order of importance according to the number of grouped trajectories in each one.
The larger a group of trajectories is, the more important/recurrent is the pattern.

The most recurrent pattern in our dataset is {\tt Pattern1} which involves 14 releases trajectories.
This pattern has the following main characteristics.
The trajectories consist of only one transition (two states), and they all end by atomic states related to bug (B) or improvement (I) issues. They start by more complex states sharing in almost all cases the aforementioned atomic states B and I, such as the states BI, BIF and BIFT.
This pattern reflects that these states are strongly connected, at least in our sample.
A simpler case of {\tt Pattern1} is {\tt Pattern3} which involves 7 trajectories, that are all consisting of only one state. This pattern represents releases which are about bug fixing or performing improvements (but not both), as in the ending state of {\tt Pattern1}.

The second recurrent pattern is {\tt Pattern5} which involves 10 releases trajectories.
All these trajectories consist of 3 transitions (4 distinct states) and, as in {\tt Pattern1}, they all end by atomic states related to bug (B) or improvement (I) issues.
This pattern is different than the other ones mainly with regard to its second state, where in almost all involved trajectories, the second state is around bug and improvement issues that are mixed with technical issues (BIT), most frequently,  or with new feature issues (BIF).
The same observation can be reported to the first state in the pattern, but here either both types of issues, new feature issues and technical issues, are cohabiting together with bug and improvement issues (the most complex state BIFT), or non of them is involved, and then the state is BI.

The main differences in {\tt Pattern4} and {\tt Pattern6} is that these patterns involve releases that all start by complex states. This is generally true during the first two states, when the software state is related to at least three types of issues, that are bug, improvement and new feature issues.
These releases actually play the role of ``issues collectors'', as they inherit a complex mixture of issues that were opened before their inception dates and take in charge the cleaning of inherited state.
In {\tt Pattern4}, the aforementioned three types of issues cohabit together after several transitions, in which the development path is mainly about sorting out technical issues and then opening new technical issues, to be sorted out next.

The final pattern that we identified is {\tt Pattern2} which involves 7 releases trajectories.
In this pattern, all trajectories consist of 3 transitions, and almost all of them are only around bug fixing and performing improvements.
The main distinguishable property of this pattern is that all involved trajectories traverse clean/Zen states (no opened issues) before facing new identified bugs, that have to be resolved during the development iterations of concerned releases.

\section{Related Work}\label{sec:relatedWork} 
The most related work to our proposal is around analyzing and documenting software releases. 
In this direction, Hindle \etal \cite{Hind07} analyze the behavior of projects around the time of release. 
They use four classes to categorize software revisions in the source control system: source code, testing, build and documentation. 
A revision belong to one of the aforementioned categories based on the files associated to the revision.    
Then, they use these classes to identify behaviors that occur around release time (\ie before or after releasing a new version), that they call as release patterns. 
In a recent work, Moreno \etal \cite{More14} propose an automated approach for generating detailed release notes, ARENA. 
ARENA relies on the code changes occurred in the commits performed between two releases. 
It summarizes those changes and link them to other information extracted from commits' notes and from the description of issues associated to the previous release. 
Our present work is complementary to the work in \cite{More14} as it can enrich the release notes with new kind of reports about the tracking process of issues. 
From another perspective, Khomh \etal \cite{Khomh12} investigate empirically whether rapid short  release cycles improve the quality of the released software, comparing the crash rates and the proportion of post-release bugs of the rapidly released versions with those having a
traditional (long) release cycle. The study shows that the number of post-release bugs remain comparable with both types of release cycle, but with shorter release cycles bugs
are fixed faster.

Another body of related work is around analyzing software evolution based on change logs for understanding the evolution process and identifying common evolution phases. 
In this research direction, 
Xing~and~Stroulia~\cite{Xing:2004:1357808} present an approach for
understanding evolution phases using structural differencing algorithm to analyze changing class
models over time and build the system's evolution profile. The resulting
sequence of changes is then analyzed to gain insight about the
system's evolution. 
Barry~et~al.\cite{Barry2003} propose a method to identify volatility-based patterns of software evolution. Volatility is approximated by computing the amplitude, dispersion
and periodicity of software changes at regular interval in the software history.
Each period is defined by a volatility class, and sequence analysis is applied
to reveal similar pattern in time. 
Kemerer \etal \cite{Kemerer1999} use also sequence analysis for mining patterns of software evolution from change logs. In their approach, change events are categorized into 30 categories that are based on (1) three subjective classifications for changes (maintenance tasks: corrections, adaptations, and enhancements); 
(2) six basic categories for changed modules (data handling, logic, structure, computation, user interface, and module interface); 
(3) and three types of modifications that can be performed in ``enhancement'' events (add, change or delete). To categorize events, only the text/comments describing the change logs are used.
In the same vein, Hindle \etal \cite{Hind09a} propose an approach to classify large commits into 5 categories of maintenance tasks:  corrective, adaptive, perfective, feature addition, and non-functional improvement.  
However, Kothari \etal \cite{Koth06} argue that it is not possible to categorize software changes in a confident and automated fashion using the aforementioned specific categories of maintenance tasks.  
They propose an automated approach that examines the temporal evolution of source code to identify {\it Change Clusters}. These change clusters classify code change activities as either a software maintenance or a new development. 
To identify change clusters, their approach fist identifies a subset of canonical changes that best represent the modification activities in a given time period. Then, using clustering techniques, it classifies all other changes in the underlined time period as being similar to one of the identified canonical clusters. The authors use change clusters to identify trends in the development process.  

In a different direction, D'Ambros~and~Lanza~\cite{Dambros:2006:1602374} propose a visualization to analyze and characterize the evolution of software entities, at different granularity levels. 
Besides, there is a family of research papers that focus on 
analyzing software change logs for identifying commits that contain tangled changes \cite{Kiri14} or peripheral modifications \cite{Kusu13}, 
providing further insights on the nature of commits \cite{Hatt08b, Drag11},
or for identifying the commit window of a release \cite{Shob14}.  

\vspace{-0.1cm}
\section{Conclusion}\label{sec:conclusion}
\vspace{-0.1cm}
We present a sequence-analysis approach to model and analyze the evolution of software states through a release, namely the release trajectory.
The states are identified based on the recorded issues in the ITS, and the release's trajectory is based on the tracking process of these issues.
We consider the recorded issues as concrete diagnostics of the software state, and we analyze the evolution of their statuses during the life-cycle of releases.
We also adapted our approach to model the tracking of software issues in the software change logs as development-based, commits-based, releases' trajectory.
Based on our model, we used sequence analysis techniques to summarize release trajectories and address important questions regarding the co-habitation of unresolved issues and the transition probability between different software states, and for mining recurrent patterns of release trajectories.
The study opens perspectives for further detailed investigations of development paths of releases and possible relations between the properties of release trajectories and other properties that can be associated to software releases, such as the cost and productivity.

\bibliographystyle{IEEEtran}
\bibliography{ReleaseTrajectory-informal}

\end{document}